\documentclass[a4paper,fleqn,usenatbib]{mnras}

\usepackage{newtxtext,newtxmath}

\usepackage[T1]{fontenc}
\usepackage{ae,aecompl}

\usepackage{graphicx}	
\usepackage{amsmath}	
\usepackage{amssymb}	
\usepackage{multirow}
\usepackage{ulem}		

\newcommand{\beq}{\begin{equation}}
\newcommand{\eeq}{\end{equation}}
\newcommand{\bea}{\begin{eqnarray}}
\newcommand{\eea}{\end{eqnarray}}
\newcommand{\bei}{\begin{itemize}}
\newcommand{\eei}{\end{itemize}}
\newcommand{\bee}{\begin{enumerate}}
\newcommand{\eee}{\end{enumerate}}
\newcommand{\noi}{\noindent}

\newcommand{\msun}{M_{\odot}}

\title{Core collapse supernovae as Cosmic Ray sources}

\author[Marcowith, Dwarkadas, Renaud, Tatischeff, Giacinti]{
Alexandre Marcowith,$^{1}$ \thanks{E-mail: Alexandre.Marcowith@umontpellier.fr}
Vikram Dwarkadas,$^{2}$
Matthieu Renaud,$^{1}$
Vincent Tatischeff,$^{3}$ 
\newauthor
and Gwenael Giacinti$^{4}$
\vspace{0.4cm} \\
$^{1}$Laboratoire Univers et Particules de Montpellier (LUPM) Universit{\'e} Montpellier, CNRS/IN2P3, CC72, place Eug{\`e}ne Bataillon,\\ 34095, Montpellier Cedex 5, France.\\
$^{2}$Department of Astronomy and Astrophysics, University of Chicago, 5640 S Ellis Ave, Chicago, IL 60637, USA.\\
$^{3}$Centre de Sciences Nucl\'eaires et de Sciences de la Mati\`ere, IN2P3-CNRS and Universit\'e Paris-Sud, F-91405 Orsay Cedex, France.\\
$^{4}$Max-Planck-Institut fur Kernphysik, P.O. Box 103980, D 69029 Heidelberg, Germany.
}

\date{Accepted XXX. Received YYY; in original form ZZZ}

\pubyear{2018}
\hypersetup{draft}
\begin{document}
\label{firstpage}
\maketitle

\begin{abstract}
Core collapse supernovae (CCSNe) produce fast shocks which pervade the dense circum-stellar medium (CSM) of the stellar progenitor. Cosmic rays (CRs) if accelerated at these shocks can induce the growth of electromagnetic fluctuations in the foreshock medium. In this study, using a self-similar description of the shock evolution, we calculate the growth timescales of CR-driven instabilities. We select a sample of nearby core collapse radio supernova of type II and Ib/Ic. From radio data we infer the parameters which enter in the calculation of the instability growth times. We find that extended IIb SNe shocks can trigger fast intra day instabilities, strong magnetic field amplification and CR acceleration. In particular, the non-resonant streaming instability can contribute to about 50\% of the magnetic field intensity deduced from radio data. This results in the acceleration of CRs in the range 1-10 PeV within a few days after the shock breakout. In order to produce strong magnetic field amplification and CR acceleration a fast shocks pervading a dense CSM is necessary. In that aspect IIn supernov\ae~are also good candidates. But a detailed modeling of the blast wave dynamics coupled with particle acceleration is mandatory for this class of object before providing any firm conclusions. Finally, we find that the trans-relativistic object SN 2009bb even if it produces more modest magnetic field amplification can accelerate CRs up to 2-3 PeV within 20 days after the outburst.
\end{abstract}

\begin{keywords}
Acceleration of particles -- shock waves -- cosmic rays.
\end{keywords}



\section{Introduction}
High-energy cosmic rays (CRs) are likely accelerated in fast shocks produced in very energetic events \citep{Bell78}. CRs above an energy of $10^{17}-10^{18}$ eV are expected to arise from extragalactic sources. Below this energy the sources are thought to be Galactic. However, few Galactic sources meet the energy confinement limit constraint $E_{\rm conf}= Ze~R~B$ \citep{Hillas84} where the particle Larmor radius $R_{\rm L}$ for a particle of charge $Ze$ in a magnetic field of strength $B$ equals the size of the source $R$. If the magnetic field strength is typical of interstellar medium (ISM) values, i.e.  $B \sim 3~\mu{\rm G}$ then the source size has to be large, at least $\sim 10$ pc for protons, in order for them to remain confined at energies up to $\sim 10^{17}$ eV. This could be the case in superbubbles \citep{Bykov01, Parizot04}. High-energy CRs can also be confined in sources with higher magnetic fields. This seems to be the case in young supernova remnants (SNRs). Many historical SNRs show thin X-ray filaments (see \citet{Parizot06} and references therein). The width of the X-ray filaments allows one to deduce a lower limit on the post-shock (downstream) magnetic field. Some objects (Cassiopeia A, Tycho, Kepler) have magnetic field strengths larger than 100 $\mu$G, much larger than the field that could be generated by a simple compression at the SNR forward shock of the ISM magnetic field.\\

The process of amplification of the magnetic field is unknown,
although many theoretical ideas have been proposed. One argument is
that TeV electrons which produce non-thermal X-rays are accelerated at
the shock front with a magnetic field amplified by the acceleration of
high-energy ions (mainly protons; \citet{Park15}). 
Magnetic field amplification (MFA) then originates from plasma instabilities 
driven by CR ions \citep{Bell04, Pelletier06, Marcowith06, Zira08, Zira208, Amato09, Bykov11} a possibility that has been further tested numerically \citep{Reville08, Riquelme10, Reville13, Caprioli14, Bai15, vanMarle18}. It appears that while MFA is
observed in a wide range of cases, final amplitudes of the magnetic
field remain uncertain, mainly because of the limited-time dynamics of
simulations that prevent them from reproducing the long timescales on
which particle acceleration evolves. An important argument raised in
\cite{Bell04} is that the fastest instability, induced by CR current
streaming ahead the shock front, has a growth rate $\Gamma_{\rm g}
\propto n_0^{1/2} V_{\rm sh}^3$ where $n_{\rm 0}$ and $V_{\rm sh}$ are
the ambient gas density and the shock velocity, respectively. Hence,
the largest magnetic field fluctuation growth rates produced by
energetic particles at an energy $E$ are obtained in dense
environments pervaded by fast shocks. Some authors \citep{Schure13,
  Marcowith14, Cardillo15} have therefore pointed to the earliest
stages of SN evolution (within months to years of explosion) as
possible PeVatron accelerators. At the time of this study the only
available data on related high-energy gamma-ray emission is an 
upper limit from the very young SNR SN 1987A obtained with the H.E.S.S. Observatory \citep{Hess15}.\\

One possibility would be to search for gamma-ray emission at a very
early expansion stage, when the forward shock is interacting with a
very dense circumstellar medium (CSM). This is the case for
core-collapse SNe, which evolve in the winds of their progenitor
stars. For a constant mass-loss rate and wind velocity, the density of
the surrounding CSM decreases as $R^{-2}$, and is thus highest at
radii close to the star. If the wind parameters are not constant, the
density decrease could be parameterized as $R^{-s}$, and the slope
differs from the value of 2. The higher the stellar mass-loss rate, and
lower the wind velocity, the higher will this density be at a given
radius. \\

Recent theoretical efforts in this area concentrate on
trans-relativistic SNe, super-luminous SNe and especially SNIIn, which are objects where either shock speeds exceed 0.1c or shocks pervade very dense CSM resulting from strong mass-loss rates $\sim
10^{-3/-2}~M_{\odot}/\rm{yr}$ \citep{Chakraborti11}. After the shock
breakout, the forward shock traveling in the CSM becomes collisionless
and CR may be accelerated efficiently, producing broadband non-thermal
emission and high-energy neutrinos \citep{Murase11, Katz12, Ellison13,
  Murase14, Zira16}, see however \citet{Giacinti15} for an alternative
scenario. \citet{Budnik08} and \citet{Ellison13} pointed out that the maximum acceleration efficiency likely occurs in the trans-relativistic regime for in particular Ib/Ic SNe with shock Lorentz factors $\beta_{\rm sh} \gamma_{\rm sh} > 1$.  GeV gamma-rays and neutrinos appear to be the best opportunities to test particle acceleration and CR production in SNe \citep{Murase14}. However GeV photons associated with interaction-powered SNe have not been detected in a Fermi-LAT data search of a sample of 147 SNe of type IIn and Ibn, with the closest being at a distance of $\sim$ 50 Mpc \citep{Ackermann15}. A search of 45 super-luminous supernovae (SLSNe) with the Fermi-LAT telescope \citep{Renaultetal17} also did not find any excess $\gamma$-rays at the SLSN positions. The same conclusion has been found for TeV photons using the H.E.S.S. observatory
\citep{Simoni17}.\\

A proper evaluation of the gamma-ray and related
multi-wavelength/multi-messenger emission during the early phase of
blast wave expansion is the main purpose of this series of papers. In
this first paper we investigate particle acceleration efficiency at
the supernova blast wave shock evolving into a dense CSM. Following
the approach adopted in \citet{Dwarkadas13} we derive a general
formalism including SN dynamics and wind properties, which can be
applied to any SN type where self-similar solutions
\citep{Chevalier82} are applicable. In this paper we assume that the
self-similar solutions are applicable even when some of the SN energy
is expended in accelerating particles. This is a reasonable assumption
  provided that the CR pressure does not exceed $\sim$ 10\% of
  the gas pressure \citep{Chevalier83,Kang10}. For the young SNe
considered in this work, this would most likely be the case, as shown
for SN 1993J \citep{Tatischeff09}. \\

The main hypothesis driving our study is that the previously invoked
CR-driven plasma instabilities are at the origin of the magnetic field
strength deduced from radio monitoring of SNe (see \citet{Marcowith14}
for a preliminary discussion and \citet{Bykov18}). Starting from this assumption we adapt
the theory of diffusive shock acceleration \citep[DSA,][]{Drury83,
  Berezhko99} to the case of fast moving forward shocks expanding into
the CSM produced by the wind of massive star SN progenitor. Within the
adopted formalism we discuss the different instabilities that may lead
to MFA and test CR acceleration efficiency at the forward shock for
different types of core-collapse SNe. We also include an accurate
treatment of the evolution of the CR maximum energy with time. However it should be noted that the origin of the magnetic field deduced from radio observations remains elusive. It can either result from a dynamo process which develops at the contact
discontinuity separating the shocked ejecta from the shocked CSM medium. The decelerating contact discontinuity can become Rayleigh Taylor unstable, producing fingers of ejecta that stretch into the shocked CSM \citep{Jun96,Bjornsson17} .\\

The results obtained herein are quite general, and applicable to any
core-collapse SN whose ejecta density profile can be described by a
power law, and which propagates in a medium whose density also
decreases as a power-law, such that the resultant shock wave can be
described by a self-similar solution. Specific calculations are made
for the case of SN 1993J, mainly because the parameters for this SN
are well known from extensive observations, and it has therefore often
been used as a testbed for radio and gamma-ray observations (see \S
\ref{S:93J}). \\

The lay-out of this article is as follows. Section \ref{S:93J}
describes the properties of SN 1993J, which is the fiducial object in
this study. Section \ref{S:CRA} details our model of shock dynamics
and CSM, including wind density profile and wind magnetic field
strengths. In particular section \ref{S:SAM} describes shock and
magnetic field dynamics of a sample of radio SNe selected on the basis
of the quality of their radio data. Section \ref{S:DSA} presents the model of particle acceleration and magnetic field amplification in SN shock waves. The maximum energy reached by CR particles is studied in Section \ref{S:MAX}. The main results of this work are discussed in Section~\ref{S:DIS} and a conclusion is finally given in Section~\ref{S:CON}.

\section{SN 1993J}\label{S:93J}

Supernova 1993J was discovered in a spiral arm of the galaxy M81
\citep{Ripero93} at a distance of 3.63 Mpc \citep{Freedman01}. It
subsequently became the optically brightest SN in the northern
hemisphere, and one of the brightest radio SNe ever detected. It
resulted from the explosion of a massive star in a binary system with
a progenitor mass ranging in the interval 13-20 $M_{\odot}$ \citep{Maund04}. The
star then evolved into a red super-giant (RSG) phase with a mass loss
rate of $\sim 10^{-6}$ to $10^{-5} M_{\odot} \rm{~yr}^{-1}$ and a slow wind $V_{w}
\sim 10~\rm{km/s}$ \citep[see][]{Tatischeff09}. \\

SN 1993J is the best monitored SN at radio wavelengths
\citep{Bietenholz10, Marti11}. It is of particular interest to test
particle acceleration and gamma-ray radiation in fast
shocks. \cite{Kirk95} developed a model of DSA for a shock propagating
in the dense CSM of SN 1987A and SN 1993J. For SN 1993J the authors
considered the case of a non-stationary wind producing a radial dependence of the CSM density $\propto R^{-3/2}$ \citep{VanDyk94}. In that case the gamma-ray flux produced by p-p interaction is enhanced and \citet{Kirk95} predicted a peak gamma-ray
flux $F_{\rm \gamma}(> 1 \rm{~TeV}) \sim 2 \times 10^{-12}~\rm{ph~cm^{-2}~s^{-1}}$. This first calculation appeared to be an overestimation because (1) subsequent calculations \citep{Fransson98} showed that a constant mass-loss profile resulted
in dynamics that were more consistent with radio measurements and
(2) the mass-loss rate assumed was higher than in later publications. \\

\citet{Tatischeff09} (T09 hereafter) used the non-linear DSA model of
\citet{Berezhko99} coupled with the self-similar hydrodynamics
expansion solutions of \citet{Chevalier82} to calculate the radio
synchrotron emission produced at the blast wave. He found that the
magnetic field was strongly amplified in the blast wave region shortly
after the explosion. Adopting the non-resonant streaming instability
\citep{Bell04} as the main process for MFA, T09 found an upstream
magnetic field strength in the shock precursor of $B_{\rm u} \sim
50(t/1~\rm{day})^{-1}$ G for a shock propagating in a wind with
constant mass-loss properties, and therefore a density profile $\rho_0
\propto R^{-2}$. T09 also found that during the first $\sim$ 8.5 years
after the explosion, about 19\% of the total energy processed by the
forward shock was used up in accelerating CRs. In this model, maximum
CR energies are quickly reached only 2 days after the SN outburst with
a peak energy $E_{\rm max} \sim 20$ PeV for protons. Finally,
accounting for the absorption of gamma-rays by the soft photons from
the SN photosphere, using an isotropic gamma-gamma opacity, T09 found
a peak gamma-ray flux above 1 TeV of $F_{\rm \gamma}(> 1 \rm{~TeV})
\sim 4 \times 10^{-15} \rm{~ph~cm^{-2}~s^{-1}}$ about 270 days after
the explosion, too low to be detected by the current Cherenkov
telescope facilities. However, in the GeV domain gamma-gamma
absorption is almost negligible, and T09 derived a peak flux $F_{\rm
  \gamma}(> 1 \rm{~GeV}) \sim 2\times 10^{-9} \rm{~ph~cm^{-2}~s^{-1}}$,
still more than one order of magnitude below Fermi-LAT sensitivity. 

\cite{Dwarkadas13} developed a general formalism to derive the
gamma-ray flux including self-similar type ejecta and wind
profiles. For the case of SN 1993J, he derived an unabsorbed gamma-ray
flux one order of magnitude above the flux
given by T09 at day 1 after outburst, but decreasing somewhat
faster, as $t^{-1.17}$ rather than $t^{-1}$. \cite{Dwarkadas13} also confirmed that actual GeV/TeV flux levels are not detectable by any active gamma-ray facilities.

\section{Shock dynamics and progenitor wind properties}\label{S:CRA}

\subsection{Shock dynamics}
\label{S:DYN}
In this study, shock radius and velocity are assumed to evolve as a
power-law with time. The initial time after the SN outburst is $t_0$
and the corresponding shock radius is $R_0$. We have: \beq
\label{Eq:RSH}
R_{\rm sh}(t)=R_0 \times \left({t \over t_0}\right)^{m} \ , 
\eeq 

and

\beq
\label{Eq:VSH} 
V_{\rm sh}(t)={R_0 m \over t_0} \times \left({t \over t_0}\right)^{m-1} \ , 
\eeq 

We note $V_0 =R_0 m / t_0$. \\

This formalism can be generalized to the case where the ejecta and
surrounding medium are power-laws, and a self-similar solution can be
used to describe the evolution \citep{Chevalier82}.  If we write the
ejecta density of the SN as ${\rho}_{ej} = A t^{-3} {\rm v}^{-k}$, and the
surrounding medium density as ${\rho}_{cs} = C\,r^{-s}$, then a
self-similar solution for the evolution of the forward shock can be
written as \citep{Chevalier94}:

\beq
\label{eq:racss}
R_{\rm sh}(t)= \beta \,\left({\frac{\alpha A}{C}}\right)^{1/(k-s)} \; t^{(k-3)/(k-s)} \ ,
\eeq

\noindent
where $\alpha$ is a constant given in \citet{Chevalier82}, $\beta$ is
the ratio of the forward shock to the contact discontinuity radius
$R_{sh}/R_{CD}$, and $m$, $k$ and $s$ are linked by $m=(k-3)/(k-s)$. The
value of $k$ is inferred for different progenitors by
\citet{Matzner99}. It is close to 10.2 for RSGs, while it is assumed
to be smaller for more compact Wolf-Rayet (WR) stars. The value of $s$
needs to be inferred for each SN from the observations.  Often, the
value of $s$ can be deduced from the X-ray light curves
\citep{Dwarkadas12}. Type IIn SNe have values of $s > 2$ at late times
after about 3 years, while IIP SNe may have $s \approx 2$.

If we consider the time in days $t_{\rm d}$ we can write this as
\beq
\label{eq:racss2}
R_{\rm sh}(t)= (86400)^{(k-3)/(k-s)} \; \beta \,\left({\frac{\alpha A}{C}}\right)^{1/(k-s)} \; t_d^{(k-3)/(k-s)} \ .
\eeq
\noindent
By comparing Eqs \ref{Eq:RSH} and \ref{eq:racss2} we can write:

\beq
\label{eq:r0} 
R_0 = (86400)^m \; \beta\,\left({\frac{\alpha A}{C}}\right)^{1/(k-s)} \ .
\eeq

\noindent
For SN 1993J, $s = 2$ and $m = 0.83$ (T09), thus $k \sim$ 7.88. For a mass-loss rate
of 3.8 $ \times 10^{-5} \msun$ yr$^{-1}$ and a wind velocity of 10 km
s$^{-1}$, we have C = 1.92 $ \times 10^{14}$ g cm$^{-1}$. According to
equation 2.4 in \citet{Chevalier94}, with the explosion energy as
10$^{51}$ ergs, and the ejected mass as 2.2 $\msun$ (T09), A=$7.6
\times 10^{75}$. With $\alpha=0.15$, $\beta=1.265$
\citep{Chevalier82}, we get \footnote{For SN 1993J T09 derive $R_0$ at a time $t_0= 100$ days (see table \ref{T:SNII}). In the rest of the study, unless specified, we assume
  that the time dependence of $R$ remains valid at a time $t_0 = 1$
  day. We use extrapolated values of the shell radius $R$ and magnetic
  field strength $B$ at $t_0 = 1$ day to test acceleration of particles
  after the outburst. The extrapolated values at $t_0 =1$ day are in relatively good agreement with the values of $R$ and $B$ derived by \citet{Fransson98} at $t_0= 10$ days. We recall this point in section \ref{S:DSA}.} $R_0 \sim 3.43 \times 10^{14}$ cm and $V_0 \simeq 3.29~\times~10^9$ cm s$^{-1}$. \\
  
In the following, we neglect the dynamical impact on the circumstellar wind of the radiation emitted at shock breakout. In reality, the flash of photons from breakout accelerates the layers of the wind close to breakout radius, $R_{\rm bo}$ ($R_{\rm bo}=R_{\star}$ for an optically thin wind), to a substantial fraction of the shock velocity at breakout. See, for example, \cite{ChevalierKlein79}. This reduces the size of the velocity discontinuity at the collisionless shock at early times, and thence the energy processed by this shock. However, photons are diluted as $1/R^{2}$ at $R>R_{\rm bo}$. Therefore, the impact of the radiation becomes negligible once the collisionless shock has reached a distance of only a few $R_{\rm bo}$. For optically thin winds, this occurs at $t<1$ day at most, which justifies our assumption.  

\subsection{Properties of the circumstellar medium}\label{S:CSM}
\subsubsection{Wind density profile}
The wind mass density scales as a power-law with an index $s$ which
depends on the mass-loss history of the progenitor. For a steady wind
(constant mass-loss rate and wind velocity) $s=2$. The mass density
experienced by the forward shock at a time $t$ is, using
Eq.(\ref{Eq:RSH}), 

\beq
\label{Eq:RHO} 
\rho_{\rm CSM}(t)=\rho_0~\left({R_{\rm sh}(t)\over R_0}\right)^{-s}= \rho_0~\left({t \over t_0}\right)^{-ms} \ , \eeq and \beq\label{Eq:DEN}\rho_0 ={\dot{M}(R_0) \over 4\pi V_{\rm w}(R_0) R_{\rm 0}^2} \simeq 1.3m_{\rm p} n_{\rm H, 0} \ , 
\eeq 

\noindent
where the factor 1.3 accounts for the presence of a medium containing
90\% H and 10\% He, and $m_{\rm p}$ and $n_{\rm H,0}$ are the proton
mass and hydrogen density at $t_0$.\\ Numerically we have for the CSM mass density at a radius $R_0$
\[
\rho_{0} \simeq \left[{5.0 \times 10^{13}\over R_{\rm{0}}^2}~\rm{g/cm^{3}}\right] ~\dot{M}_{-5}(R_0)~V_{\rm w,10}(R_0)^{-1} \ ,
\]
where, the shock radius at $t_0$ is expressed in cm, the progenitor mass-loss rate $ \dot{M}$ is expressed in units of
$10^{-5} M_{\odot} \rm{~yr}^{-1}$ and the wind asymptotic speed $V_{\rm w}$ is in units of 10 km/s. The mass-loss rate is derived at a fix radius $R_{\rm ref} = 10^{15}$ cm (see the discussion in \citet{Fransson96}).  The mass-loss rate at $R_0$ is by definition given by $\dot{M}(R_0)=\dot{M}(R_{\rm ref})\left(R_{\rm ref}/R_0\right)^{2-s}$. \\
We consider the wind velocity to be constant with the radius. This assumption is justified as soon as min($R_0,R_{\rm ref}$) is larger than the stellar radius $R_\star$. The main mechanism which drives RSG winds is not known yet: radiation pressure on dust grains, effect of magneto-acoustic waves, turbulent pressure due to convection and radiative pressure on molecular lines may contribute to mass ejection in these objects \citep{Josselin07, Haubois09, Auriere10}. The wind velocity evolution in RSG is therefore uncertain. \citet{Josselin07} use a tomography technique to probe line velocity profiles in the atmosphere of a sample of RSG. They find speeds in the range 10-30 km/s. We assume that beyond a few stellar radii, the wind is accelerated to its velocity at infinity. Hence, if min$(R_0, R_{\rm ref}) > R_\star$ then $V_{\rm w}(R_0) = V_{\rm w}(R_{\rm ref})$. Unless otherwise specified we choose $R_\star \simeq 10^3 R_\odot \simeq 6.96~10^{13}~\rm{cm}$ for RSG stellar radius. In the case of SN 1993J we adopt $\dot{M}_{-5}(R_{\rm ref}) \simeq 3.8$ and $V_{\rm w,10}(R_{\rm ref}) =  1$ (see the discussion in T09). Both mass-loss and wind velocity are taken identical at $R_0$ as $R_0 \sim 5 R_\star$ and $R_{\rm ref} \sim 14 R_\star$ and because we set s=2. \\
WR winds can be described using a CAK \citep{Castor75} profile for a radiation-driven wind, namely $V_{\rm w}(r)=V_{\rm w}(\infty)\left(1-R_\star/R\right)^b$. The index b is not well-constrained. \citet{Hillier03} invokes b values in the range 1 to 3 but also argues that WR winds have a structure different from O star radiation driven winds, which implies that a single value of b can not reproduce the wind velocity profile properly. \citet{Nugis02} propose a model for the optically thick part of the wind where b depends on the WR type (WN or WC) and lies in a range between 2.9 and 6.5. \citet{Vink11} find lower values for b in the range 1.5-2. Since WR have lost their outer envelopes and are more compact than RSG stars, the stellar radius $R_\star$ is much smaller than $R_{\rm ref}$ and $R_0$, we identify $V_{\rm w}(R_{\rm Ref})$ and the wind velocity at infinity $V_{\rm w}(\infty)$ and we assume $V_{\rm w}(R_0) = V_{\rm w}(R_{\rm Ref})$.\\

The mass-loss properties of SN 1993J are not shared with the entire class of Type IIb SNe. \citet{Chevalier10} have suggested that there exist two sub-classes of Type IIb SNe, those with extended (eIIb) and those with compact (cIIb) progenitors. SN 1993J falls in the former category. The ones in the latter category may arise from compact progenitors such as WR stars. Their mass-loss rates are lower, and their wind velocities could be significantly higher by up to two orders of magnitude, leading to wind densities that are almost two orders of magnitude lower than that of SN 1993J. An example is SN 2008ax, which had a mass-loss rate a few to 10 times lower \citep{Chornock11, Roming09}, and whose spectrum bears similarity to Ib SNe, which are expected to arise from WR stars.
More generally speaking, the wind properties of SN progenitor stars vary considerably with SN types. Mass-loss rates of massive stars are described in \citet{Smith14}. Type IIP SNe are associated with RSG progenitors, and their wind velocities are similar to those of the type IIb SNe, lying between 5-20 km s$^{-1}$. One would expect their mass-loss rates to also span the total range of mass-loss rates of RSGs, from 10$^{-7}$ to 10$^{-4} M_{\odot} ~\rm{yr^{-1}}$ \citep{Mauron11}. However their X-ray emission indicates that their mass-loss rates lie on the lower end of the
range, and do not appear to exceed 10$^{-5} M_{\odot} ~\rm{yr^{-1}}$
\citep{Dwarkadas14}. Type Ib/c SNe are thought to arise from WR stars,
whose radiatively driven winds have velocities of 1000-3000 km
s$^{-1}$, and mass-loss rates ranging from $5 \times 10^{-7}$ to
$5 \times 10^{-5} M_{\odot} ~\rm{yr^{-1}}$. The most diverse class is
the type IIn SNe, which have the highest optical and X-ray luminosities, indicative of high mass-loss rates. A mass-loss rate of 10$^{-3} M_{\odot} ~\rm{yr^{-1}}$ was noted for SN 2005kd \citep{Dwarkadas16}, and a rate as high as 10$^{-1} M_{\odot} ~\rm{yr^{-1}}$ has been found for SN 2010jl \citep{Fransson14}. Such high rates are not easily explained by stellar winds of RSG or WR stars, but are perhaps characteristic of eruptive outbursts from luminous blue variable (LBV) stars. Not all type IIn SNe have such high mass-loss rates however. Fitting the radio light curves of SN 1995n suggested a mass-loss rate of $6 \times 10^{-5} M_{\odot} ~\rm{yr^{-1}}$ \citep{Chandra09}, whereas hydrodynamical and X-ray modeling of SN 1996cr indicated an even lower mass-loss rate $< 10^{-6} M_{\odot} ~\rm{yr^{-1}}$. The wind velocities of type IIn SNe are not well calibrated either. A velocity of $\sim$ 100 km s$^{-1}$ is deduced from the narrow component of spectral lines in SN 2010jl. The expansion velocity of the CSM was found to be $\sim$ 90 km s$^{-1}$ for SN 1997ab \citep{Salamanca98}, $\sim$ 45 km s$^{-1}$ for SN 1998S \citep{Fassia01}, $\sim$160 km s$^{-1}$ for SN 1997eg \citep{Salamanca02} and $\sim$ 100 km s$^{-1}$ for SN 2002ic \citep{Kotak04}. This suggests that velocities of 100 $\pm$ 50 km s$^{-1}$ for the winds of Type IIn SNe are common, but it should be emphasized that there could be much wider variation in these velocities. Mass-loss rates and wind velocities can also vary over time, sometime episodically, as often seems to happen near the end of a star's life \citep{Foley07, Margutti14}. Type IIn SNe progenitors winds hence show complex structures which are difficult to properly account in a self-similar model. This statement should be kept in mind while considering some of these objects in section \ref{S:SAM}.\\

\subsubsection{Wind magnetic field} Magnetic field strength and topology in massive  star winds are difficult to measure. \citet{Walder12} review magnetic fields on the surface of massive stars and in their winds. Fields can be deduced from maser polarization observations using different type of tracers probing different media around the star. At growing distance from the star SiO$_2$, H$_2$O and OH masers are used successively \citep{Vlemmings02}.  There is usually no clear trend on the distance dependence of measured magnetic fields in the CSM of evolved stars; profiles with $B\propto R^{-\alpha}$ with $\alpha=1-3$ can give reasonable fits to the data. \citet{Vlemmings17} perform polarisation analysis of circumstellar dust and molecular lines in the RSG star VY CMa. The authors found a polarisation consistent with a toro\"{\i}dal magnetic geometry but higher angular resolution are necessary to confirm this trend. The magnetic field strength has some uncertainty but could be as high as 1-3 G. \citet{Auriere10} obtain a longitudinal (along the line of sight) magnetic field strength of the order of 1 G in $\alpha$ Ori (Betelgeuse), the most well-studied RSG star. Gauss-level field strength has been confirmed in two other RSGs by \citet{Tessore17}. We can compare the above values to a magnetic field strength at the stellar surface obtained by a balance between magnetic field energy density and wind kinetic energy density  \citep{Fransson98, Uddoula02}\footnote{We decide to take as a reference the strength of the magnetic field in equipartition with the wind kinetic pressure rather than the thermal pressure since winds in RSG stars are cold with temperatures $T \sim 10^4$ K. \citet{VanMarle12} perform hydrodynamical simulations of RSG winds. Using their set-up parameters we find that both pressures have the same order. WR winds are hotter with temperatures $T \sim 10^5-10^6$K but have comparable mass loss and speeds that can be two orders of magnitude larger.}
\[
B_{\rm eq,0} \simeq \left[{2.5 ~ 10^{13} \over R_{\rm 0}}~\rm{G}\right]~\dot{M}_{-5}^{1/2}V_{\rm w,10}^{1/2} \ . 
\]
With stellar values appropriate for Betelgeuse, $R_0 = 10 R_\star \simeq
8.3~10^{14}$~cm, $\dot{M}_{-5} \simeq 0.3$ and $V_{\rm w,10} \simeq 1.5$ \citep{Smith09}, we get $B_{\rm eq} \simeq 0.03$ G. Hence, this value extrapolated at $R_\star$ produces a magnetic field $\ga 0.3$ G. But this extrapolation is highly sensitive to $R_0$ and to the radial dependence of $B_{\rm eq}$. An uncertainty of an order of magnitude of the wind magnetic field with respect to $B_{\rm eq}$ is assumed in this study. \\

WR stars have fast winds which produce strong line broadening and
hence make magnetic field measurements difficult. \citet{Hubrig16}
report on magnetic field measurements in a set of 5 WR stars, with
strengths for the line of sight component in the range 200-300 G. As
an example, we consider the particular object WR 6 (class WN4) in this sample, which shows a longitudinal magnetic component of $B_{\rm z} =
258\pm 78$ G. As a matter of comparison, we can evaluate the
equipartition magnetic field $B_{\rm eq,0}$ using the stellar
parameters derived from \citet{Nugis00}: $R_0 = 10 R_\star \simeq
6 \times 10^{11}$cm, $\dot{M}_{-5} \simeq 0.62$, $V_{\rm w, 10} \simeq 146$ and assuming s=2, for the wind speed profile we find $B_{\rm eq,0} \simeq 400$ G. It is however hazardous to compare directly the two values as the observations provide only a mean longitudinal magnetic field strength.\\
 
From the above considerations, we assume in this study a CSM magnetic field strength proportional to $B_{\rm eq}$ with 
\beq
\label{Eq:BWI} 
B_{\rm  w}(t) \simeq \varpi B_{\rm eq,0} \left({t  \over t_0}\right)^{-ms\over 2}
\eeq 
\noindent
where we assume the ratio $\varpi= B_{\rm w}(t_0)/B_{\rm eq,0}$ to be in the range 0.1--10. The time dependence arises from the radial dependence of the wind density as mentioned in section \ref{S:DYN}.\\
 
In Eq. (\ref{Eq:BWI}) as soon as $R(t) \gg R_\star$, the wind magnetic field scales as $1/R_{\rm sh}$ which is expected in case of a toroidal geometry. As discussed above, there is an important uncertainty on magnetic field topology and radial dependence in the wind of evolved massive stars. RSG winds also show quite inhomogeneous and turbulent structures \citep{Smith09}. Except for the case $\varpi \gg 1$, the magnetic field should also reflect such inhomogeneity and depart from a simple toroidal configuration \footnote{Accounting for another field geometry can be
  treated allowing the parameter $\varpi$ to be dependent on the
  distance to the star surface. This would require at
  minimum two more parameters to be introduced, both of which are poorly constrained: the value of
  $\varpi$ at $R_0$ and a dependence of $\varpi$ with
  $R$, the spectral index, which would generally be a power-law. In
  order to keep our formulation as simple as possible we assume a
  constant equipartition parameter $\varpi$ in the rest of the
  study.}. \\ 
  
The ambient Alfv\'en velocity $V_{\rm A,CSM} = B_{\rm W}/\sqrt{4\pi \rho_{\rm CSM}} = \varpi V_{\rm w}$ and the CSM
magnetization ${\cal M}= (V_{\rm A,CSM}/c)^2$ is
\[
{\cal M}\simeq \left[1.1~10^{-9}\right] \varpi^2  V_{\rm w,10}^2\ .
\] 
Considering $\varpi$ to be in the range 0.1-10 we always obtain ${\cal M} \ll 1$ whatever the type of progenitor.

\subsubsection{Ionization of the Circum-Stellar Medium}\label{S:ION}  Another important parameter entering in the calculation of particle acceleration efficiency is the degree of ionization of the pre-shock
medium. Neutrals have a strong impact over magnetic fluctuations which could raise around the shock front \citep{Drury96, Reville07}. \\

The ionization of the medium is a complex problem, since it depends
not only on the progenitor star but on the presence of any nearby
companions, or its location in an association or cluster of stars,
which may also serve to ionize the medium.\\

While the SN explosion itself tends to ionize the medium, the
densities close to the stellar surface are so high that recombination
occurs quickly. As shown for example in \citet{Dwarkadas14}, for the
mass-loss rates and velocity assumed for the wind medium around SN
1993J, the medium around the star that the shock traverses in the
first few months can be considered to have recombined. However the
X-ray emission from the SN itself can ionize the medium. This depends
on the quantity $\chi = L_x / (nR^2)$, where $L_x$ is the X-ray
luminosity, $n$ is the density and $R$ is the radius from the shock
\citep{Kallman82}.  \citet{Dwarkadas16} writes this in terms of the
mass-loss rate and velocity.  For a steady wind with constant
mass-loss parameters, this reads as:
\beq
\chi = 2 \times 10^{-38} \, L_x \,{\xi}^{-2}\left[\frac{\dot{M}_{-5}}{V_{w,10}}\right]^{-1} \;\;,
\label{eq:ion}
\eeq

\noindent
where $\xi = [1 + 2n(He)/n(H)]/[1 + 4 n(He)/n(H)] \sim 0.9$. If we use
the X-ray luminosities given in \citet{Chandra209} we find that $\chi
> 100$ for at least 11 days but less than 20 days. $\chi > 100$ is
required for ionization of intermediate elements like C, N, and O;
ionization of heavier elements like Fe requires $\chi > 1000$.\\

However this is an underestimate, because these luminosities are only
in the 0.3-8 keV band. The X-ray emission early on has much higher
temperatures, so the luminosity in this range is a small fraction of
the total X-ray luminosity. In the case of SN 1993J, \citet{Leising94}
and \citet{Fransson96} indicate that the total luminosity at 12 days
was about 5.5 $\times$ 10$^{40}$ erg s$^{-1}$ in the 50-150 keV range, and even at day 28.5 it was 3.0 $\times$ 10$^{40}$ erg s$^{-1}$. This would suggest that the intermediate elements were ionized at least for the first month, and presumably longer. It is unlikely however that the medium was fully ionized except possibly in the first week - heavy elements such as Fe and Ni would presumably be partially but not fully ionized after a week.\\

An added complication is the presence of the companion star to SN
1993J. \citet{Fox14} have shown that this is likely a B2 star, and as
such may provide some additional UV ionizing flux. However without
more accurate details regarding its distance and observing surface
temperature, it is difficult to estimate its effect.\\

Eq. \ref{eq:ion} can be written for a more general CSM density profile as \citep{Dwarkadas14}:

\beq
\chi = 2 \, L_{x_{38}} \,{\xi}^{-2}\left[\frac{\dot{M}_{-5}}{{V_{w,10}}}\right]^{-1}\,V_4^{s-2}\left[\frac
{t_{\rm{d}}}{8.9}\right]^{s-2} \;\;.
\label{eq:ion2}
\eeq
\noindent
where $L_x{_{38}}$ is the X-ray luminosity in units of 10$^{38}$ erg
s$^{-1}$, $V_4$ is the maximum ejecta velocity scaled to 10$^4$ km
s$^{-1}$, $t_{\rm d}$ is the time in days. Since the X-ray luminosities of even the most luminous SNe
are at most 10$^{42}$ erg s$^{-1}$, we have $L_{x_{38}} \la 10^4$. The ionization parameter then depends on the slope of the density
profile. For $s < 2$, the last two terms are raised to a
negative power and the ionization parameter decreases with time,
whereas for $s > 2$ it is increasing with time. However the other
quantities are generally decreasing with time. It is clear that the
medium will rarely be fully ionized outside of the first few weeks at
best, although partial ionization is likely for many SNe for a couple
of years.\\

Neutrals can contribute to partially quench the growth of CR-driven instabilities \citep{Reville07}. From the above discussion the ionization fraction $X=n_{\rm i}/n_{\rm tot}$ defined as the ratio of ion to total medium densities is likely close to 1 at least during the weeks after the explosion because light elements are fully ionized by the blast wave X-rays. Hence, CR-driven instabilities should be weakly corrected with respect to the fully ionized solution \citep{Reville07}. However, a description of a long yearly term evolution of the shock environment and CR-driven instabilities would require more accurate modeling, which is beyond the scope of this work, but will
be explored in future. 

\subsection{Properties of a selected sample of SNe}\label{S:SAM}
Today about 200 SNe have been detected at radio wavelengths, but only a few are sufficiently close to show well-resolved light curves. A first list of type II SNe detected by VLBI includes: SN 1979C (SN
IIL), SN 1986J (SN IIn), SN 1987A (SN IIpec), SN 1993J (SN IIb),
SN1996cr (SN IIn), SN 2008iz (possibly SNIIb see \citet{Mattila13}), and SN 2011dh (SN IIb) (\citet{Bartel17} and references therein) \footnote{The core-collapse SN type is associated with the SN name when available.}. In this list we discard SN 1987A and SN 1996cr:
the former is interacting with a high density HII region and
shell close in to the shock \citep{Blondin93, Chevalier95, Dewey12},
which cannot be reproduced in a self-similar model; while the latter
has too scarce data at early epochs to carry out a proper evaluation
of shock dynamics and magnetic fields \citep{Bauer08}. We add to
this list SN 2001gd (SN IIb) \citep{Perez05, Stockdale07} detected
by VLBI, VLA and GMRT facilities. We select a list of type Ib/c SNe
based on the same data quality criterion: SN 1983N (SN Ib), SN 1994I
(SN Ic), SN 2003L (SN Ic) \citep{Weiler86, Soderberg05}. We also
include SN 2009bb a relativistic Ibc SN \citep{Soderberg10}. \\

The modeling of the radio lightcurves at different wavebands requires
accounting for a number of processes: synchrotron self-absorption
(SSA), free-free absorption (FFA) by the ambient thermal plasma
(internal FFA) or by CSM matter (external FFA), and plasma processes
like the Razin-Tsytovich effect \citep[see][]{Fransson98}, although the last one has generally not been found to be important in supernov\ae. Radio
observations are important for many aspects of particle acceleration
modeling. First, the spectral turnover produced by SSA leads to an
estimate of the magnetic field intensity of the synchrotron emitting
zone. Second, the synchrotron spectral index provides a constraint on
the electron distribution function and then on the acceleration
process. Third, radio images are used to derive the SN shell dynamics
and the time evolution of the shock radius and velocity, which are
mandatory for any microphysical calculations of particle acceleration
efficiency (see T09). The shell radius and speed can also be compared
to a self-similar expansion model \citep[see section \ref{S:DYN}
  and][]{Chevalier82}.\\

The properties of the radio emission depend on the SN type: Type Ib/c
SNe show steep spectral indices ($\alpha > 1$, with a radio flux
scaling as $S_{\nu} \propto \nu^{-\alpha}$), and have similar 
radio luminosity peaking before optical maximum at a wavelength around 6~cm, while Type
II SNe show flatter spectra ($\alpha < 1$) with a wider range of radio
luminosities usually peaking at 6~cm significantly after optical
maximum \citep{Weiler02}. We summarize in Tables 1 and
2 the main properties of the above SNe (when available) that can be deduced from VLBI observations. Apart from $\alpha$, we use the following notations: the magnetic field strength is characterized by its amplitude $B_0$ at a reference time $t_0$ to be specified and by its time dependence given by $B(t) = B_0 (t/t_0)^{-n}$. The VLBI shell radius is characterized by its radius at $t_0$, $R_0$ and by the index $m >0$ (see Eq.\ref{Eq:RSH}). We also show the deduced progenitor mass-loss rate in units of $\rm{10^{-5} M_\odot/yr}$. To derive this last parameter it is usually necessary to make an assumption on the progenitor wind speed. This is also specified in both tables.


\begin{table*}\label{T:SNII}
\begin{tabular}{|c | c| c| c| c| c| c|}
\hline 
SN name  & $\alpha$ & $t_0(\rm{days})$  & $(R_0(\rm{cm}), m)$  & $\dot{M}(\rm{10^{-5} M_\odot/yr})$ & $V_{\rm w}(\rm{10~km/s})$ & $(B_0(\rm{G}), n)$ \\
\hline
SN 1979C & $0.74^{+0.05}_{-0.08}$& 5  & $(8.7(e14), 0.91\pm0.09)$ & 12  & 1 & $([20-30], -1.00)$\\
\hline
SN 1986J & $0.67^{+0.04}_{-0.08}$ & 5 & $(3.2(e15), 0.69\pm0.03)$ & 4-10 & 1  & $([30-50], -1.00)$ \\
\hline
\multirow{2}{*}{SN 1993J} & 1.00 & 10 & $(1.9(e15), 1.00)$  & 5  & 1 & $(25.5, -0.93\pm0.08)$  \\ 
					& 0.90 & 100  & $(1.6(e16), 0.829\pm0.005)$ & 3.8 & 1 & $(2.4 \pm 1.0, -1.16\pm0.20)$ \\
\hline
{SN 2001gd} & $1.0\pm0.1$ & 500 & $(3.6(e16), 0.845)$ & 2-12 & 1 & (0.05-0.35,$-$) \\
\hline
SN 2008iz & 1.00 & 36  & $(8.8(e15), 0.86\pm0.02)$ & 3.7 & 1 & $([0.4-3.2], -1.00)$\\
\hline
\multirow{2}{*}{SN 2011dh}  & 1.15 & 4  & $(5.0(e14), 1.14\pm0.24)$ & $0.01$ & $1$ & $(5.9, -1.00\pm0.12)$\\ 
					  & 0.95 & 15 & $(3.1(e15), 0.87\pm0.07)$  & 0.07 & 2 & $(1.1, -1.00)$ \\ 
\hline
\end{tabular}
\caption{Magnetic field and shock radius evolution for a set of type
  II SNe. The following references have been used for the different
  sources in the list: SN 1979C \citep{Weiler91, Marcaide09, Marti211,
    Lundqvist88}, SN 1986J \citep{Weiler90, Bietenholz210, Marti211},
  SN 1993J upper row \citep{Fransson98} and lower row (T09), SN2001gd
  \citep{Perez05}, SN
  2008iz \citep{Kimani16, Marchili10}, SN 2011dh upper row
  \citep{Horesh13} and lower row \citep{Krauss12, Yadav16}. The spectral index $\alpha$ corresponds to the optically thin
  synchrotron spectrum. {\bf SN 1979C} and {\bf SN 1986J}: The magnetic field strength
  derived by \citet{Marti211} is obtained at $t_0=5$ days from a synchrotron model . Upper and lower values for $B_0$ are associated with the uncertainties of $\alpha$ and m. The value of $n$ is obtained from the solution of propagation in a wind with constant mass-loss rate with $s=2$. The radius $R_0$ at day 5 is extrapolated from angular size measurements obtained by VLBI observations. 
  {\bf SN 1993J}: \citet{Fransson98} use an expansion index m=0.74 at
  $t > 100$ days, \citet{Marti11} present a long term radio survey
  where the expansion law index at frequencies above 1.7 GHz varies
  from $m_< = 0.925 \pm 0.016$ before $t_{\rm br}= 360\pm 50$ days to
  $m_>=0.87\pm 0.02$ after. As the index is not specified at timescales close to the breakout we use m=1. {\bf SN2001gd}: The magnetic field strength
  derived by \citet{Perez05} in the same manner as in \citet{Marti11}. Mass-moss rates values are consistent with the results presented in \citet{Stockdale07} where a value for s=1.61, i.e. different from a steady mass loss, has been used. {\bf SN2008iz}: \citet{Kimani16} derive the
  equipartition magnetic field strength, the upper and lower limits
  depend on whether protons are accounted for in the estimation or
  not. {\bf SN 2011dh}: We keep the two expansion parameter solutions derived by \citet{Horesh13} and \citet{Krauss12} respectively. Note that \citet{deWitt16} reported on a value of the expansion parameter $m = 0.91 \pm 0.01$ compatible with the Horesh et al solution.}
\end{table*}

\begin{table*}\label{T:SNI}
\begin{tabular}{|c| c| c| c| c| c| c|}
\hline 
SN name  & $\alpha$ & $t_0(\rm{days})$  & $(R_0(\rm{cm}), m)$  & $\dot{M}(\rm{10^{-5} M_\odot/yr})$ & $V_{\rm w}(\rm{1000~km/s})$ & $(B_0(\rm{G}), n)$ \\
\hline
SN 1983N & $1.03\pm0.06 $ & 13  & $(2.3(e15), 0.81)$ & 50 & 1 & $(3.6, -1.00?)$ \\
\hline
SN 1994I & $1.22 $ & 10.125  & $(2.4(e15), 1.00)$ & 3  & 1 & $(2.28, -)$ \\
\hline
SN 2003L  & $1.1$ &10  & $(4.3(e15), 0.96)$ & 0.75 & 1 & $(4.5, -1.00)$\\
\hline
SN 2009bb & $1.0$ &20  & $(4.4(e16), 1.00)$ & $0.2\pm0.02$& 1 ?& $(0.6, -1.00)$ \\
\hline
\end{tabular}
\caption{Magnetic field and shock radius evolution for a set of type
  Ib/Ic SNe, adapted from the following references: SN 1983N
  \citep{Sramek84, Weiler86, Slysh92}, SN 1994I \citep{Weiler11,
    Alexander15}, SN 2003L \citep{Soderberg05}, SN 2009bb
  \citep{Soderberg10, Chakraborti11}. {\bf SN 1983N}:  the
  magnetic field strength dependence with time is not available, and
  is therefore assumed to be $\propto t^{-1}$.  \citet{Slysh92}
  only derives estimates of an upstream magnetic field strength of 0.9
  G which is multiplied by 4 to account for shock
  compression. \citet{Sramek84} derive a mass-loss rate of $\sim
  5 \times 10^{-6}~\rm{M_\odot/yr}$ with $V_{\rm w} =10$ km/s but decide to fix $V_{\rm w}=1000$ km/s, a value 
  which seems more reasonable for this type of SN. {\bf SN 1994I}:
  the fitted parameters of the model derived by \citet{Alexander15} do
  not show a simple power-law time dependence for both $B$ and
  $R$. \citet{Alexander15} invoke two extreme mass-loss rates
  depending on the progenitor terminal wind speed. \citet{Weiler11}
  deduce from radio observations a mass-loss rate of $\dot{M} \sim
  2 \times 10^{-7}~\rm{M_\odot/yr}$ for $V_{\rm w} =10$
  km/s. We again select the solution corresponding to a wind velocity $V_{\rm w} = 1000$ km/s. \citet{Alexander15} use an expansion index m=0.88. However, \citet{Bjornsson17} criticized this assumption deduced from a self-similar solution. We assume by default this parameter to be 1.0 (corresponding to the free expansion solution). {\bf SN 2003L}: the solution corresponds to model 1 of \citet{Soderberg05}. The authors provide two other models with faster shock deceleration. Their model 2 also includes a shallower CSM density profile with $\rho \propto r^{-1.6}$. {\bf SN 2009bb}: \citet{Soderberg10} do not specify the terminal wind speed
  assumed in order to calculate the mass-loss rate. We take $V_{\rm w}
  =1000$ km/s by default.}\label{T:SNI}
\end{table*}

\section{Diffusive shock acceleration and magnetic field amplification}\label{S:DSA}
This section describes the model of particle acceleration and magnetic
field amplification after the SN outburst. \\

In sections \ref{S:GEN} and \ref{S:MFA} we apply the model to SN~1993J, using the $R$ and $B$ parameters derived by T09, but extrapolated at $t_0 = 1$ day.  It should be kept in mind that these extrapolated values may not be precise, as T09 and
\citet{Fransson98} only derived them at $t_0 = 100$ days and $10$ days
respectively, from the modeling of radio light curves  (see also the
footnote of section \ref{S:DYN}).

\subsection{Acceleration models}\label{S:GEN}
We adopt a model for particle acceleration at collisionless shocks
based on the theory of DSA \citep{Drury83}. The highest CRs have an
upstream diffusion coefficient $\kappa_{\rm u}$ which fixes the length
scale of the CR precursor $\ell_{\rm u}= \kappa_{\rm u}/V_{\rm
  sh}$. The timescale to advect the frozen CR-magnetized fluid to the
shock front is 
\beq\label{Eq:ADV} 
T_{\rm adv,u}= {\kappa_{\rm u} \over V_{\rm sh}^2} \ . 
\eeq 
CRs at energies close to $E_{\rm max}$ stream ahead of the shock
and simultaneously generate electromagnetic fluctuations. The upstream
diffusion coefficient at these energies can be expressed with respect
to the diffusion coefficients parallel and perpendicular to the
background wind magnetic field. This coefficient depends on two
parameters \citep{Jokipii87}: $\eta$, the ratio of the parallel CR
  mean free path to CR Larmor radius $R_{\rm L}$, and $\theta_{\rm B}$
  the magnetic field obliquity. We write the parallel diffusion
coefficient as $\kappa_{\parallel} = \eta R_{\rm L} v/3$, where $v$ 
is the particle speed. The minimum value $\eta=1$ corresponds to the Bohm diffusion limit. In parallel shocks ($\theta_{\rm B}$ =0)
$\kappa_{\rm u} = \kappa_{\parallel}$ while in perpendicular shocks
($\theta_{\rm B}= \pi/2$) it matches the perpendicular diffusion coefficient, i.e. $\kappa_{\rm u} = \kappa_{\perp}$. Without
considering magnetic field line wandering in the wind turbulent medium
we have $ \kappa_{\perp}=\kappa_{\parallel}/(1+\eta^2)$. Hence, if
$\eta \gg 1$ diffusion is suppressed in the perpendicular shock
case. If the magnetic field in the wind is purely toroidal and weakly
turbulent the advection timescale $T_{\rm adv, u}$ drops. On the other
hand, if the wind medium has some level of turbulence (see the
discussion in section \ref{S:CSM}) then we can expect to have a
diffusion coefficient close to Bohm ($\eta \sim 1$), and to have a
non negligible portion of the shock in the parallel configuration.\\

We define as model P and model T the two extreme configurations described above. In model P the wind magnetic field is assumed to 
be parallel. The advection time is in this case
\beq
\label{Eq:ADVP} 
T_{\rm adv,u,P}\simeq {\eta_{\rm P} R_{\rm L} v \over 3 V_{\rm sh}^2} \ .  
\eeq 
We account for some turbulence in the wind medium and include a
contribution due to perturbations in the wind magnetic field, $\delta B_{\rm u}$, which is assumed to be in equipartition with the mean field strength $B_{\rm w,0}$: $B_{\rm w}^2 = \delta B_{\rm u}^2 + B_{\rm w,0}^2$.  This turbulence is assumed to be injected at large wind scales, typically the wind termination shock radius, and $\delta B_{\rm u} \simeq B_{\rm w,0}$ at the highest CR energies.

Using Eq. (\ref{Eq:BWI}) for the wind mean magnetic field and expressing the proton Larmor radius $R_{\rm L} \simeq E/ e B_{\rm w}$ for a particle energy of 1 PeV as
\beq
\label{Eq:RGW} 
R_{\rm L} \simeq 3.3~10^{12}E_{PeV} B_{\rm w,G}^{-1}~\rm{cm}  \ ,
\eeq 
we find an advection time (expressed in seconds) 
\bea
\label{Eq:TAD} 
T_{\rm adv,u,P} &\simeq& \left[{1.3~10^9 \eta_{\rm P} R_{\rm 0,cm} \over V_{\rm 0, cm/s}^2 \varpi}~\rm{s} \right] \times \nonumber \\ 
&& {E_{\rm PeV} \over \dot{M}_{-5}^{1/2}  V_{w,10}^{1/2}}~\left({t \over t_0}\right)^{2(1-m)+m{s\over2}} \ .  
\eea 

In model T the wind mean magnetic field is assumed to be toroidal and weakly perturbed
with fluctuations of strength $\delta B_{\rm u,w} < B_{\rm w, 0} \simeq B_{\rm w}$.  The advection time is in this case
\beq
\label{Eq:ADVT}
T_{\rm adv,u,T} \simeq {R_{\rm L} v \over 3 \eta_{\rm T} V_{\rm sh}^2}\ . 
\eeq 

For the parameters adopted for SN 1993J we have $T_{\rm adv,u,T}
\simeq (0.24~\rm{day})\times (1/\eta_{\rm T}\varpi) E_{\rm PeV}
t_{\rm d}^{1.17}$ and $T_{\rm adv,u,P} \simeq (0.24~\rm{day})\times
(\eta_{\rm P}/\varpi) E_{\rm PeV} t_{\rm d}^{1.17}$, hereafter $t_{\rm d}$ is the time in units of days after the SN explosion. \\ 

We can deduce the acceleration timescale from the above estimates 
\beq
\label{Eq:TAC} 
T_{\rm acc, P}=g(r) T_{\rm adv,u, P}  = g(r) {\kappa_u \over V_{\rm sh}^2}
\eeq
\noi
where $g(r)=3r/(r-1) \times (1+ \kappa_d r/\kappa_u)$ depends on the
shock compression ratio $r$ and on the ratio of the downstream to
upstream diffusion coefficients. The ratio $\kappa_{\rm d}/\kappa_{\rm
  u}$ depends on the magnetic field
obliquity and on the shock compression ratio $r$. We have
$\kappa_{\rm d}/\kappa_{\rm u} = r_{\rm B}^{-1}$ with $r_{\rm B} = B_{\rm d}/B_{\rm u}$ is the ratio of magnetic fields in the postshock region and in the wind and $g(r)=3r/(r-1)\times (1+r/r_{\rm B})$. In the model P, we have $r_{\rm B} \simeq 1$ and $g(r)=3r(r+1)/(r-1)$. In the model T the magnetic field is weakly perturbed and perpendicular to the shock normal and $\kappa_{\rm d}/\kappa_{\rm u} = r^{-1}$ and $g(r)=6r/(r-1)$. \\

The calculations in this section have not assumed {\it any}
magnetic field amplification (MFA). MFA at the shock precursor by the
streaming of high-energy CRs produces a reduction of the precursor
length. In the case of Bohm diffusion the reduction factor corresponds
to the ratio of the amplified to ambient magnetic field
strengths. If the perturbations in the precursor are isotropic then $r_{\rm B} \simeq \sqrt{(1+2r^2)/3}$ \citep{Parizot06} and $g(r)$ is modified accordingly.

\subsection{Magnetic field amplification}\label{S:MFA}
In this section we discuss different CR driven instabilities that may operate at the SN forward shock and generate magnetic field fluctuations necessary for the DSA process to operate at a high efficiency. All calculations are performed in the framework of model P but we discuss the case of particle acceleration in model T in \S \ref{S:MOT}. \\

In this work we do not consider fluid instabilities triggered by the CR pressure gradient in the precursor \citep{Drury12}. Fluid perturbations generate magnetic field fluctuations through a small-scale dynamo process \citep{Beresnyak09}. The magnetic field growth can be fast if the velocity stretching in the CR precursor is strong enough. This happens when the shock is modified by the CR pressure. We did not consider this possibility in this work and assumed that the shock modification is weak (see T09). We postpone to a future study the case of strongly CR-modified shocks and their impact over the different instabilities which may grow in the CR precursor. 

\subsubsection{Bell non-resonant streaming instability}\label{S:BEL}
\citet{Bell04} discussed a process of magnetic field amplification in
SNRs driven by currents produced by the streaming of CRs ahead of the
shock front. The CR streaming induces a return current in the
background plasma, which triggers magnetic fluctuations at scales $\ell
\ll R_{\rm L}$, where $R_{\rm L}$ is the Larmor radius of the CRs
producing the current. This instability is non-resonant and can be
treated using a modified MHD model \citep{Bell04, Bell05, Pelletier06}. 
 
\paragraph{Growth timescale} \citet{Bell04} gives the minimum growth timescale 
(corresponding to the maximal growth rate) for the non-resonant
streaming (NRS) instability
\beq
\label{Eq:TN0} 
T_{\rm min,NRS}={2 \phi \over \xi_{\rm CR}} ~{R_{\rm L} c \over V_{\rm sh}^3} V_{\rm A, CSM} \ , 
\eeq 
where $\phi \simeq \ln(E_{\rm max}/m_{\rm p} c^2)$ is fixed by the
maximum CR energy. This expression
implicitly assumes that the CR distribution at the shock scales with
the particle momentum as $p^{-4}$. In other words the CR distribution
is close to the test-particle solution and a softer distribution scaling as $p^{-k}$ with $k > 4$ would lead to $\phi \simeq 1/(k-4)$. T09 finds that this is a
reasonable assumption for the case of SN 1993J and we assume that
  this condition holds true in this study. A more detailed modeling including the effects of CR back-reaction is beyond the scope of this paper, but will be addressed in future work. In Eq.(\ref{Eq:TN0}), we take $\xi_{\rm CR}$ to be the fraction of the shock ram pressure imparted to CRs. T09 finds $\xi_{\rm CR} = \xi_{\rm CR0} \times (t/t_0)^{1-m} \propto 1/V_{\rm sh}$. This scaling is in a strict sense only valid as long as CR back-reaction over shock dynamics is weak. Then, we can write $\xi_{\rm CR} \propto p_{\rm inj}/V_{\rm sh}^2$, where $p_{\rm inj}$ is the CR injection momentum. It varies as $p_{\rm th}$, the momentum of thermal shocked plasma, itself being proportional to $V_{\rm sh}$
\citep{Blasi05}. Once $P_{\rm CR}$ is beyond a certain fraction of shock ram pressure, the acceleration process becomes non-linear and CR escaping upstream carry an increasing energy flux which back-reacts over the injection process and hence, this scaling should be revised and the injection process becomes less efficient. The inclusion of this non-linear physics is beyond the scope of this simple study and will be addressed in a forthcoming work. \\

Using Eq.(\ref{Eq:BWI}) the NRS growth rate reads (using $V_{\rm A,CSM} = \varpi V_{\rm w}$)
\bea
\label{Eq:TNR}
T_{\rm min,NRS} &\simeq& \left[{2.2 ~10^{18} R_0 \over V_0^3}~\rm{s} \right] ~ {\phi_{14} \over \xi_{\rm CR,0.05}} E_{\rm PeV} \times \nonumber \\
&& \dot{M}_{-5}^{-1/2}~V_{w,10}^{1/2}~\left({t \over t_0}\right)^{2(1-m)+{ms \over 2}}.
\eea 
We assume $E_{\rm max} \simeq 10^{15}$ eV and we note $\phi_{14}=\phi/14$. 

\paragraph{Advection Constraint} A necessary condition for the NRS instability
to grow is ${\cal R}_{\rm NRS}=T_{\rm min,NRS}/T_{\rm adv,u} < 1$ hence
using Eqs. (\ref{Eq:TNR}) this condition reads
\beq
{\cal R}_{\rm NRS} \simeq \left[1.7~10^9 \varpi \over \eta V_0\right]~{\phi_{14} \over \xi_{\rm CR0,0.05}} V_{\rm w,10}  < 1 \ , 
\eeq
which is independent of time because $\xi_{\rm CR}$ scales as
$1/V_{\rm sh}$. In the model P we have $\eta \sim 1$ (we drop the subscript p hereafter). In the case of
SN 1993J this gives ${\cal R}_{\rm NRS} \simeq 0.5 \varpi
\phi_{14}/\xi_{\rm CR0,0.05}$.  The equipartition parameter $\varpi$
is important, we see that if $\varpi > 2$, the NRS instability can
not grow. The most favorable condition for the instability to grow is
the case of sub-equipartition wind magnetic field.\\

A useful quantity is the wavenumber corresponding to the maximum NRS
instability growth rate. It is given by
\beq
k_{\rm max}^{-1} = T_{\rm min,NRS} V_{\rm A,CSM} \ .
\eeq
Using Eq.(\ref{Eq:TNR}) we express it in cm units
\bea
\label{Eq:KMA}
k_{\rm max}^{-1} &\simeq& \left[{2.2~10^{24} R_0 \over V_0^3}~\rm{cm} \right]~{\phi_{14} \varpi \over \xi_{\rm CR0,0.05}} E_{\rm PeV} \times \nonumber \\
&&\dot{M}_{-5}^{-1/2} \times V_{w,10}^{1/2}~\left({t \over t_0}\right)^{2(1-m)+m{s\over 2}} ,
\eea
which for SN 1993J is
\[k_{\rm max}^{-1} \simeq 2.1~10^{10}~\rm{cm}~{\phi_{14} \varpi \over \xi_{\rm CR0,0.05}} E_{\rm PeV}~t_{\rm d}^{1.17},\] where again $t_{\rm d}$ is the time in day units.\\

In \citet{Pelletier06} the saturation magnetic field is (see their Eq.28)
\beq
B_{\rm sat,NRS}^2=12\pi {\xi_{\rm CR} \over \phi} \rho_{\rm CSM} {V_{\rm sh}^3 \over c} \ .
\eeq
Hence we have
\bea\label{Eq:BNR}
B_{\rm sat,NRS} &\simeq&  \left[224 {V_0^3 \over R_0^2}~\rm{G}\right]^{1/2}~\sqrt{{\xi_{CR0,0.05} \over \phi_{14}}} \times \nonumber \\
&& \dot{M}_{-5}^{1/2}~V_{w,10}^{-1/2}~\left({t \over t_0}\right)^{m-1-{ms\over 2}} \ .
\eea

For $s=2$, $B_{\rm sat,NRS}$ varies as $t^{-1}$. A typical value of the
NRS saturation field in the case of SN 1993J is $B_{\rm sat,NRS} \simeq
16~\rm{G} \sqrt{\xi_{CR0,0.05}/ \phi_{14}} \times t_{\rm d}^{-1}$, a
result already obtained in T09.\\

Using Eqs.(\ref{Eq:BWI}) and (\ref{Eq:BNR}) we define the
amplification factor, the ratio of the saturation and wind magnetic
fields:
\bea
\label{Eq:CAA}
{\cal A} = {B_{\rm sat,NRS} \over B_{w}}& \simeq &\left[ {6.0~10^{-13} V_0^{3/2} \over \varpi}\right]~{1 \over V_{w,10}} \times \nonumber \\
&& \sqrt{{\xi_{CR0,0.05} \over \phi_{14}}}~\left({t \over t_0}\right)^{m-1} \ .
\eea

For SN 1993J we obtain an amplification factor ${\cal A} \simeq
113/\varpi \times \sqrt{\xi_{CR0,0.05}/\phi_{14}}\times
t_{\rm d}^{-0.17}$, slowly decreasing with time.\\

In sections \ref{S:RES}, \ref{S:FIL} and \ref{S:LON} we calculate the growth rate of instabilities which produce long wavelength perturbations. These perturbations are necessary to confine high energy CRs around the shock front.  


\subsubsection{Resonant streaming instability}\label{S:RES}
The streaming of CRs faster than the local Alfv\'en speed is known to produce long-wavelength modes at scales $\ell \sim R_{\rm L}$. To derive the instability growth rate we follow the treatment presented in
\citet{Amato09}. We introduce the following parameter $\sigma = 3 \xi_{\rm CR} V_{\rm sh}^3/\phi c$. 
We have
\beq
\sigma  \simeq [3.5 \times 10^{-13} V_0^3]~{\xi_{\rm CR0,0.05} \over \phi_{14}}~\left({t \over t_0}\right)^{2(m-1)}~\rm{cm^2/s^2} \nonumber
\eeq
and 
\beq
{\sigma \over V_{\rm A,CSM}^2} \simeq [3.5~10^{-25} V_0^3]~{\xi_{\rm CR0,0.05} \over \varpi^2  V_{\rm w,10}^2 \phi_{14}}~ \left({t \over t_0}\right)^{2(m-1)} \ . \nonumber
\eeq
We note in passing that the NRS mode exists only in a particular wave number interval $k\in [k_1,k_2]$ with $k_1 R_{\rm L}= \sqrt{\pi}/2 \sqrt{\sigma/V_{\rm A,CSM}^2}$ and $k_2 R_{\rm L}= \sigma/V_{\rm A,CSM}^2$. It disappears if $\sigma/V_{\rm A,CSM}^2 =\pi/4$ \citep{Amato09}.\\

For long wavelengths (with wave numbers such that $k R_{\rm L} < 1$), the minimum resonant streaming (RS) growth rate in the non-linear regime is given by (see \citet{Amato09}, their Eq. 40)
\beq
T_{\rm min,RS} \simeq R_{\rm L }\sqrt{{8 \over \pi \sigma}} \ . 
\eeq
We find using Eq. (\ref{Eq:BWI}) 
\bea\label{Eq:TRE}
T_{\rm min,RS} &\simeq& \left[{3.6~10^5 R_0 \over \varpi V_0^{3/2}}~\rm{s}\right] \times \nonumber \\
&&{E_{\rm PeV} \over \dot{M}_{-5}^{1/2} V_{\rm w,10}^{1/2}}~\left({\phi_{14} \over \xi_{\rm CR0,0.05}}\right)^{1/2}~ \left({t \over t_0}\right)^{1-m+m{s\over 2}} \ .
\eea
Similar to the NRS instability case, we define the ratio of the RS growth time to the advection time (Eq. \ref{Eq:TAD}) 
\beq
{\cal R}_{\rm RS} \simeq \left[{2.7~10^{-4} \sqrt{V_0} \over \eta}\right]~\left({\phi_{14} \over \xi_{\rm CR0,0.05}}\right)^{1/2}~\left({t \over t_0}\right)^{(m-1)} \ .
\eeq 
For SN 1993J this ratio is $ \sim 15/ \eta \sqrt{\phi_{14}/\xi_{\rm CR0,0.05}}~t_{\rm d}^{-0.17}$ but we note that it decreases with time, and it may rapidly be lower than 1 if $\eta$ is larger than a few. \\

\citet{Bell01} (their Eq.14 in regime A \footnote{This regime is valid for $\xi_{\rm CR} \lesssim 12 V_{\rm sh}/c$. This regime is verified for the fast shocks which develop in the SN context.}) give the level of saturation of the magnetic field if only the resonant mode is destabilized:
\beq
B_{\rm sat,RS}={B_{\rm W} \over 2}  \xi_{\rm CR} {V_{\rm sh} \over V_{\rm A,CSM}} \ .
\eeq
or using Eqs.(\ref{Eq:RHO}) and (\ref{Eq:DEN})
\beq
B_{\rm sat,RS} \simeq \left[{6.3 \times 10^5 V_0 \over R_0}~\rm{G}\right] ~ {\xi_{\rm CR0,0.05} \dot{M}_{-5}^{1/2} \over V_{\rm w,10}^{1/2}} ~ \left({t \over t_0}\right)^{-m{s\over2}} \ .
\eeq
This estimate accounts for the non-linear growth of the resonant waves.
But if the non-resonant mode is also destabilized the total magnetic saturation magnetic field is \citep{Pelletier06} (their Eq.37)
\beq
B_{\rm sat,RS} = B_{\rm sat,NRS}\left( \xi_{\rm CR} c \over V_{\rm sh} \right)^{1\over 4} \ ,
\eeq
where the multiplicative term is
\beq
 \left( \xi_{\rm CR} c \over V_{sh} \right)^{1/4} \simeq 197 ~ \left( \xi_{\rm CR0,0.05} \over V_0 \right)^{1/4} ~  \left({t \over t_0}\right)^{{(1-m)\over2}} \ . 
\eeq
Using Eq. (\ref{Eq:BNR}) we find in that case that $B_{\rm sat,RS} \propto \left({t \over t_0}\right)^{{(m-1)\over 2}-m{s\over 2}}$.

\subsubsection{Filamentation instability}\label{S:FIL} 
\citet{Reville12} demonstrate that cosmic rays form filamentary
structures in the precursors of supernova remnant shocks
due to their self-generated magnetic fields. They show that the filamentation
resulted in the growth of a long-wavelength instability,
with a minimum growth time (given by their Eq. 13)
\beq\label{Eq:TFI}
T_{\rm min,Fil}= \sqrt{\phi \over \xi_{\rm CR}} {\bar{R}_{\rm L} c \over V_{\rm sh}^2}  
\eeq
where $\bar{R}_{\rm L}$ is the CR Larmor radius taken in the amplified magnetic field produced by the NRS streaming instability. 
We note $E_{\rm th}$ 
the threshold energy to trigger the filamentation instability, i.e.
\beq\label{Eq:RGM}
\bar{R}_{\rm L,th} \ge {V_{\rm sh} \over k_{\rm max} c} \ .
\eeq
For the parameters adopted for SN 1993J, this threshold is independent of time. Using Eqs.(\ref{Eq:VSH}), (\ref{Eq:KMA}) and (\ref{Eq:BNR}) we find
\beq\label{Eq:ETH}
E_{\rm PeV, th} \simeq {331 \over \sqrt{V_0}}~\sqrt{\phi_{14} \over \xi_{\rm CR0, 0.05} } \varpi~E_{\rm PeV} \ .
\eeq
At energies $E > E_{\rm th}$, the filamentation instability is destabilized and grows with the rate given by Eq.(\ref{Eq:TFI}) 
\bea\label{Eq:TFII}
T_{\rm min,Fil}&\simeq& \left[{1.1~10^{23} R_0 \over V_0^{7/2}}~\rm{s}\right] ~\left({\phi_{14} \over \xi_{\rm CR0,0.05}}\right) \times \nonumber \\ 
&&\dot{M}_{-5}^{-1/2} V_{\rm w,10}^{1/2}~E_{\rm PeV}~\left({t \over t_0}\right)^{{5\over 2}(1-m)+m{s\over 2}} \ .
\eea
Once the NRS instability is onset and the energy threshold to trigger the filamentation instability is reached, then long-wavelength modes can be very rapidly generated by the production
of filaments. For the case of SN 1993J we find $T_{\rm min, Fil} \sim (0.1~\rm{day})\times(\phi_{14}/ \xi_{\rm CR0,0.05})~E_{\rm PeV}\times t_{\rm d}^{1.255}$.

\subsubsection{Instability generating long oblique modes}\label{S:LON}
\citet{Bykov11} show that the presence of turbulence with scales shorter than the CR gyroradius enhances the growth of modes with scales longer than the gyroradius for particular polarizations. The mode growth time is (given by their Eq.37) 
\beq
T_{\rm ob}=\sqrt{{4 \over \pi {\cal A}}} ~ \sqrt{{1\over k k_{\rm max} V_{\rm A,CSM}^2}} \ .
\eeq
The mimimum growth time is obtained for $k=1/(\eta R_{\rm L})$
\beq
T_{\rm Min,ob}=\sqrt{{4 \over \pi {\cal A}}} ~\sqrt{{\eta R_{\rm L} \over k_{\rm max} V_{\rm A,CSM}^2}} \ .
\eeq
Again we have $k_{\rm max}^{-1} = T_{\rm min,NRS} V_{\rm A,CSM}$. 
Using Eqs.(\ref{Eq:BWI}), (\ref{Eq:TNR}) and (\ref{Eq:CAA}) we find
\bea\label{Eq:TOB}
T_{\rm min,ob}&\simeq& \left[{7.9~10^{11} R_0 \sqrt{\eta} \over \sqrt{\varpi}  V_0^{9/4}}~\rm{s}\right] ~ \left({\phi_{14} \over \xi_{\rm CR0,0.05}}\right)^{3/4} \times  \nonumber \\
&& \left({1\over V_{\rm w,10}  \dot{M}_{-5}}\right)^{1/2}~E_{\rm PeV} \left({t \over t_0}\right)^{{3\over 2}(1-m)+m{s\over 2}} \ .
\eea
Using the parameters selected for SN 1993J we find $T_{\rm min,ob}\simeq (0.6~\rm{day})\times{\sqrt{\eta} \over \sqrt{\varpi}}(\phi_{14}/\xi_{\rm CR0,0.05})^{3/4}~E_{\rm PeV} \times t_{\rm d}^{1.085}$.
Following the onset of the NRS instability the oblique modes grow fast, but still with a slower rate compared to the filamentation modes. 

\subsubsection{Obliquity effects}\label{S:MOT}
In the previous calculations we have assumed that model P applies, but the orientation of the mean magnetic field with respect to the shock normal can have an impact on the growth of CR driven instabilities. We consider the case of a CR current $\vec{j}_{\rm CR}$ perpendicular to the background magnetic field $\vec{B}_{\rm w}$. The NRS instability still grows unless the modes have wavenumbers $\vec{k} \perp \vec{B}_{\rm w}$. The growth time for modes propagating parallel to the background magnetic field involves both background sound and Alfv\'en speeds \citep{Bell05}
$T_{\rm NRS,T} \simeq T_{\rm NRS,P} \times \sqrt{c_{\rm s} V_{\rm A} /2(c^2_{\rm s} + V^2_{\rm A})}$. In the meantime, the advection time given by Eq.(\ref{Eq:ADVT}) is shortened because 
the parameter $\eta > 1$. If the factor $1/(\eta \varpi) \ll 1$ then the NRS instability can be quenched in this field configuration. However, if the wind magnetic field is in sub-equipartition 
and if $1/(\eta \varpi) \sim 1$ the ratio $T_{\rm NRS}/T_{\rm adv} < 1$ and the NRS instability can grow. \\

Several recent studies have investigated the efficiency of CR injection and acceleration in high obliquity shocks with some different conclusions. \citet{Bell11} using a Vlasov model including
a collision term induced by electromagnetic process find particle acceleration at perpendicular fast supernova remnant shocks. The obliquity effects tend to sharpen the CR distribution with respect to the parallel shock case. \citet{Caprioli214} using a hybrid code including a kinetic treatment of non-thermal particles find an acceleration efficiency dropping beyond obliquity angles $\sim 45^o$ for shock Mach numbers up to 50. Protons crossing high obliquity shocks are only accelerated by the shock drift process and cannot repeatedly cross the shock front because downstream advection is faster. The typical sound speed in massive star winds is $c_{\rm s} \sim 10~\rm{km/s}~(T/10^4~\rm{K})^{1/2}$. Early in the SN evolution, shock velocities are of order $10^4$ km/s, therefore we have typical shock Mach numbers $~10^3~(T/10^4~\rm{K})^{-1/2}$. It would be interesting to test this limit angle of acceleration efficiency for such high Mach number shocks. 
\citet{vanMarle18} using a PIC module for CRs in a MHD shock solutions find that high obliquity shocks produce CR acceleration especially because the particles accelerated by the shock drift
mechanism are able to induce enough turbulence downstream do corrugate the shock front. The shock corrugation produces patterns where the local magnetic field is parallel and where particle that cross the shock from downstream can trigger NRS/filamentation instability upstream. It requires long multi-scale simulations possibly in 3D to characterize particle acceleration at highly oblique shocks and we consider the issue still opened (see also \citet{Caprioli18}). \\

Finally, another point raised by \citet{Zirakashvili18} is that the wind magnetic field is radial at close distances to the progenitor star. Then, fast shocks can inject CR efficiently during the early shock propagation phase. Once CRs are injected the onset of CR driven magnetic fluctuations are able to maintain a parallel magnetic field over some fraction of the shock front, and thereby maintain particle injection.

\subsection{CR driven wave growth in supernovae}\label{S:CRG}
\begin{figure}
\centering
\includegraphics[width=8cm]{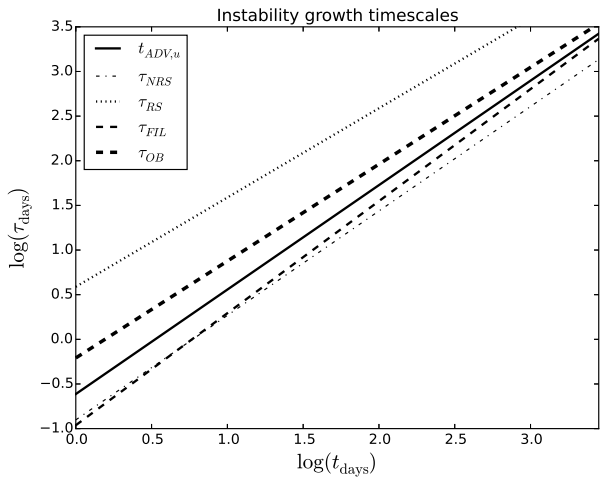}
\caption{Main instability growth timescales as a function of the time in days for the fiducial case SN 1993. We have assumed $\eta=\varpi=1$, $E=1$ PeV, $\phi=14$, $\xi_{\rm CR}=0.05$.}
\label{F:WG}
\end{figure}

We consider first the example of SN 1993. In figure \ref{F:WG} we plot the advection time and the different growth timescales for model P for particles with energies of 1 PeV. We find at every times $\cal{R_{NR}} < $1, hence NRS modes can grow. Large scale modes, for the adopted set of parameters, can be produced by the filamentation instability. The oblique mode instability and the resonant streaming instability have growth timescales larger by a factor $\sim 2.5$ and a factor $\sim 15$ than the advection time and, for this set of parameters can not grow. However, these timescales drop more rapidly with time and at some stage can become shorter than the advection time. They can hence compete with the filamentation instability to produce long-wavelength perturbations.\\

Table \ref{T:TSN} shows values at $t=t_0$ of all relevant timescales for the other SNe in our sample. In all cases the index s = 2 is selected. We can see that in almost all cases the parameter $R_{\rm NRS}$ is in the range 0.3-3. In these cases magnetic field amplification by the NRS instability seems possible given the uncertainties on the other parameters. It also appears that the saturated magnetic field strength is a fraction, typically about 10-20\% of $B_0$ displayed in tables 1 and 2. As $B_{\rm sat,NR} \propto \sqrt{\xi_{\rm CR}}$ doubling this percentage would require to convert 50\% of the kinetic power of the shock into energetic particles which is unlikely. Investigating this regime would require non-linear diffusive shock acceleration calculations, which is beyond the scope of this simple study. However, the final saturation value can (and should) be higher if long-wavelength modes are destabilized (note that we have checked that in every cases where ${\cal R_{NR}} < 1$ we have $E_{\rm th} \ll $1~PeV, so the filamentation instability can be destabilized at this energy). Finally, the transversal magnetic field component is compressed at the shock front. This means that CR driven instabilities, if they can be triggered, can contribute to at least 50\% of the magnetic field strength inferred from modeling the radio observations. Among all sources, SN 1986J, a Type IIn SN, appears to produce the fastest instability growth and produce the highest amplification factor. This effect is due to the high mass-loss rate and high shock velocity in SN 1986J, which has a $R_0$ a factor of 3.7 larger than SN 1979C at the same time $t_0=5$~d. However, the results for SN 1986J have to be considered to be optimistic. Fast mode growths are mostly due to the small size shock value at $t_0$ because the expansion index is close to the Sedov value (0.69 versus 2/3). The combination of the radio expansion and the measured VLBI radius lead to a very high value the shock speed $V_0$. The self-similar solution is questioned by the fit of the X-ray emission \citep{Dwarkadas12} as it is the case for most of type IIn SNe. This class requires a dedicated modeling of the shock dynamics and the results obtained here are subject to an important uncertainty. SN 2001gd, SN 2011dh, SN 1983N (low wind speed solution), SN 1994I (low wind speed solution) show instability growth timescales larger than $t_0$ by factors $< 10$. In these cases, the growth of the instability may be limited by the age rather than a condition ${\cal R}_{\rm NR} < 1$. 
\\ 
If the progenitor wind velocity is high, as it is the case in Type Ib/Ic SNe, the ambient gas density drops and the NRS growth rate also. For instance, one can see in table \ref{T:DRI} that SN 2003L has ${\cal R_{NR}} \sim 40$ (and $T_{\rm min, NRS} \sim 12$ days). Here, magnetic field amplification by the NRS instability is not likely because the advection towards the shock is too fast to allow for NR modes to grow unless $\varpi \ll 1$. Another point to be mentioned in fast progenitor wind SNe is that the parameter ${\cal R_R} < {\cal R_{NR}}$. In the case of SN 2003L we find ${\cal R_R} \sim 5$. To illustrate this trend in table \ref{T:DRI} we show values of $T_{\rm min, R}$, ${\cal R_R}$ and ${\cal R_{NR}}$ for type Ib/Ic SNe. In the case of SN2009bb the NRS seems to be able to amplify the magnetic field at the shock precursor (see table \ref{T:TSN}). In this object, with respect to type II SNe, the smallness of the $\dot{M}/V_{\rm w}$ ratio is compensated by a shock reaching a mildly relativistic regime.\\
Acknowledging for these results including their limitations, we confirm with our calculations that in order to amplify the magnetic field by CR streaming a fast shock pervading a dense medium is necessary.

\begin{table*}
\begin{tabular}{|c | c| c| c| c| c| c| c|}
\hline 
SN name  & $T_{\rm min,NRS}$(days) & ${\cal R_{\rm NR}}$ & $B_{\rm sat, NRS}$(Gauss) & ${\cal A}$ & $T_{\rm min, RS}$(days) & $T_{\rm min, Fil}$(days) & $T_{\rm min, ob}$(days)  \\ 
\hline
SN 1979C & 1.0 & 0.9 & 4.7 & 47.1 & 13.3 & 1.2 & 3.3 \\
\hline
SN 1986J  & 0.2 & 0.3 & 4.5 & 219.2 & 13.8 & 0.2 & 1.6  \\
\hline
\multirow{2}{*}{SN 1993J} & 2.0 & 0.8 & 1.8 & 61.9 & 34.3 & 2.2 & 7.4  \\
                          & 57.8 & 1.1 & 0.1 & 36.1 & 568.6 & 73.7 & 160.9 \\
\hline
SN 2001gd  &  992.3 & 2.4 & 0.02 & 11.2 & 3.0(e3) & 1.9(e3) & 1.5(e3) \\
\hline
SN 2008iz & 8.1 & 0.7 & 0.4 & 72.0 & 158.2 & 8.2 & 31.8 \\
\hline
\multirow{2}{*}{SN 2011dh} & 28.4 & 1.0 & 0.2 & 40.2 & 311.0 & 34.9 & 83.4\\
						   & 33.1 & 0.8 &0.1 & 57.0 &  514.3 & 36.3 & 115.8\\
\hline
\hline
SN 2009bb & 1.5 & 6.8 & 0.06 & 24.4 & 10.1 & 0.5 & 0.3 \\
\hline
\end{tabular}
\caption{Main instability timescales of our SNe sample in the case of low progenitor wind speed ($V_{\rm w} <$ 100~km/s). All values are derived with: $\eta$ = 1, $\varpi$ = 1, $\phi_{14}$ = 1, $\xi_{CR,0.05} = 1$, $E_{\rm PeV} = 1$. For SN 1986J and SN 2001gd we use the mean value of the mass-loss rate given in table \ref{T:SNII}. In the case of SN 1993J we adopt two different set of parameters derived at two different times by \citet{Fransson98} ($t_0$=10 days) and T09 ($t_0$ = 100 days).}
\label{T:TSN}
\end{table*}

\begin{table*}
\begin{tabular}{|c | c| c| c| c|}
SN name  & $T_{\rm min,RS}$(days) &${\cal R_{R}}$ &  $B_{\rm sat, RS}$(Gauss)  & ${\cal R_{NR}}$ \\
\hline
SN 1983N & 2.0 & 11.0 & 0.3 & 102.4 \\ 
\hline
SN 1994I & 4.0 & 14.1 & 0.1 & 62.0 \\
\hline
SN 2003L & 6.3 & 12.0 & 0.06 & 35.6 \\
\hline
\end{tabular}
\caption{Resonant instability timescales in high progenitor wind SNe with $V_{\rm w} > 100$ km/s, to the exception of SN 2009bb where the shortest growth timescales are due to the non-resonant instability.}
\label{T:DRI}
\end{table*}

\section{Maximum cosmic ray energies}\label{S:MAX}
The maximum CR (hadrons) energy is fixed by five different processes: the shock age limitation, the finite spatial extend of the shock, the generated current limitation, the nuclear interaction losses, and the adiabatic losses.\\

To the exception of $E_{\rm max, cur}$ due to the current limitation, all maximum energies expressions can take two different values depending if the background magnetic field in which high-energy CR gyrate is assumed to be the wind magnetic field or the field amplified by the NRS instability. In the latter scenario calculated maximum energies have to be seen as lower values, because large-scale magnetic perturbations can be destabilized either by filamentation or oblique mode dynamo generation.   

\subsection{Age-limited maximum energy}
We can write the acceleration time as $T_{\rm acc} =(1/E dE/dt)^{-1}$ and the maximum energy 
\beq
E_{\rm max}(t)-E_{\rm max}(t_0)=\int_{t_0}^t dt \times dE/dt= \int_{t_0}^t dt \times E/T_{\rm acc}(E) \ .
\eeq
The acceleration time given by Eq.(\ref{Eq:TAC}) is $T_{\rm acc}= T_{\rm adv,u} g(r)$. In the case of model P, using Eq. (\ref{Eq:TAD}) we have
\bea
E_{\rm max,age}(t) &\simeq& \left[{7.7 \times 10^{-10} V_0^2 t_0 \varpi \over \eta R_0 (1-2m+m{s\over 2}) g(r)}~\rm{PeV}\right] \times \nonumber \\
&&\dot{M}_{-5}^{1/2} V_{\rm w,10}^{1/2}~\left(1-\left({t \over t_0}\right)^{2m-1-m{s\over 2}}\right) \ , 
\eea
if $s > {2 \over m} (2m-1)$. Otherwise $E_{\rm max, age} \propto \left(\left({t \over t_0}\right)^{2m-1-m{s\over 2}}-1\right)$, and  it grows as $\ln(t/t_0)$ in the case $s = {2 \over m} (2m-1)$. \\

In the model T the maximum energy is $E_{\rm max, age, T}=E_{\rm max, age, P} \times (g_{\rm P}/g_{\rm T}) \times \eta_{\rm P} \eta_{\rm T}$.
If we consider the background magnetic field as being $B_{\rm sat,NRS}$ then the previous maximum energy must be multiplied by the amplification factor ${\cal A}$ and $E_{\rm max,age}(t) \propto \left(1-({t \over t_0})^{3m-2-ms/2}\right)$ if $s > {2 \over m} (3m-2)$ and $E_{\rm max,age}(t) \propto \left(({t \over t_0})^{3m-2-ms/2}-1\right)$ or $E_{\rm max,age}(t) \propto \ln(t/t_0)$
otherwise.

\subsection{Geometrical losses maximum energy}
Geometrical losses are given by the condition: $\kappa_{\rm u}= \eta_{\rm esc} V_{\rm sh} R_{\rm sh}$, where $\eta_{\rm esc}$ is a parameter in the range 0.1-0.3 used to mimic the effect of particle loss in spherical geometry \citep{Berezhko96}. Hence, for model P we have
\bea\label{Eq:ESC}
E_{\rm max,esc}(t)&= &[2.7\times 10^{-10} V_0~\rm{PeV}]~{\varpi \over \eta} {\eta_{\rm esc} \over 0.3} \times \nonumber \\
&&\dot{M}_{-5}^{1/2} V_{w,10}^{1/2}~\left({t \over t_0}\right)^{2m-1-m{s\over 2}} \ .
\eea
We note that if we consider the magnetic field to be given by the saturation value obtained for the NRS instability (see Eq.(\ref{Eq:BNR}) the maximum geometrical loss-limited energy is multiplied by the amplification factor ${\cal A}$ and $E_{\rm max,esc}(t) \propto  \left({t \over t_0}\right)^{3m-2-ms/2}$.

\subsection{Current driven maximum energy}
If the NRS instability operates, the maximum energy is fixed by the number ${\cal N}$ of e-folding times growth of the NRS instability, i.e. $t/T_{\rm min,NRS}={\cal N}$. We have ${\cal N} \in [1,\ln{\cal A}]$, where the lower limit ${\cal N}=1$ corresponds to the minimum time for the instability to grow in the linear phase while ${\cal N}=\ln{\cal A}$ corresponds to the amplification by a factor ${\cal A}$ of the magnetic field. Following \citet{Schure13} (their Eq.~4) the maximum CR energy is given by the relation
\beq
E_{\rm max,cur} \ln\left({E_{\rm max,cur} \over m_{\rm p} c^2}\right) ={\sqrt{\pi} \xi_{\rm CR} \over {\cal N}} \times q\sqrt{\rho} R_{\rm sh} {V_{\rm sh}^2 \over c} 
\eeq
where $E_{\rm max}$ is in erg.\\
Finally,
\bea\label{Eq:CUR}
E_{\rm max,cur}~\phi_{14} &\simeq& \left[{4.5 \times 10^{-19} V_0^2\over {\cal N}}~\rm{PeV}\right]~\xi_{\rm CR0,0.05} \times  \nonumber \\ 
&&\dot{M}_{-5}^{1/2} V_{\rm w,10}^{-1/2} \times  \left({t \over t_0}\right)^{2m-ms/2-1}
\eea

\subsection{Maximum energy from the nuclear interaction losses}
High-energy CRs interact with ambient matter through p-p interaction with a cross-section given by \citep{Kafexhiu14} 
\bea
\sigma_{\rm pp} &\simeq& 30~\rm{mb}\left(1.89+0.18\ln(E_{\rm PeV})+6\times10^{-3}\ln^2(E_{\rm PeV})\right) \nonumber \\
&&\times \left(1-{4\times10^{-13}\over E_{\rm PeV}^{1.9}}\right)^3 = 30~\rm{mb}~\bar{\sigma}_{\rm pp}(E_{\rm PeV}) \ ,
 \eea
 and a loss timescale $T_{\rm pp} \simeq \left(K_{\rm pp} \sigma_{\rm pp} \langle n_{\rm H}(t)\rangle ~c\right)^{-1}$ with  $K_{\rm pp} \sim 0.2$. The hydrogen density $n_{\rm H}$ in the wind is obtained from Eq.(\ref{Eq:DEN}), and we account for the residence times of CRs upstream and downstream of the shock $T_{\rm u/d} = 4\kappa_{\rm u/d}/(V_{\rm u/d} c)$, where $V_{\rm u/d}$ are the upstream and downstream fluid speeds in the shock rest frame. Then the mean density experienced by a CR during a Fermi cycle is
\[
\langle n_{\rm H}\rangle = {n_{\rm H} T_{\rm u} + n_{\rm d} T_{\rm d} \over T_{\rm u} + T_{\rm d}}
\]
We use $n_{\rm d} = 4 n_{\rm H}$ in the case of a weakly modified shock (see discussion in section \ref{S:MFA}).
We then have $\langle n_{\rm H} \rangle = 4F n_{\rm H}$ with 
\[
F={1+{r_{\rm B}\over 4r} \over 1+r_{\rm B}/r} 
\]
and $n_{\rm H}$ is given by Eq.(\ref{Eq:DEN}). Finally,
\beq\label{Eq:TPP}
T_{\rm pp} \simeq \left[{6.0\times10^{-23}~R_0^2 \over \bar{\sigma}_{\rm pp}(E_{\rm PeV})}~\rm{s} \right] \times \left({ V_{\rm w,10} \over F \dot{M}_{-5}}\right) \times  \left({t \over t_0}\right)^{ms} \ .
\eeq
Comparing this time with $T_{\rm acc}$ given by Eq.(\ref{Eq:TAC}) we obtain 
\bea\label{Eq:NUC}
E_{\rm max, nuc} &\simeq& \left[{4.8\times10^{-32}~R_0~V_0^2 \varpi \over g(r) F \eta\bar{\sigma}_{\rm pp}(E_{\rm PeV}) }~\rm{PeV}\right] \times \nonumber \\
& & \nonumber \\
& &\dot{M}_{-5}^{-1/2} V_{\rm w,10}^{3/2} \times \left({t \over t_0}\right)^{2(m-1)+m{s\over2}} \ .
\eea

\subsection{Maximum energy from the adiabatic losses}
Due to the rapid flow expansion CRs also suffer from adiabatic losses. To account for the residence of CRs upstream and downstream of the shock we use Eq.1 of \citet{Volk88} and find
\beq
T_{\rm Ad} \simeq \left[{6 R_0 \over V_0}{r\over 4(r-1)}~\rm{s}\right]\left(1+{r_{\rm B}\over r}\right)~\left({t \over t_0}\right)
\eeq
In the case of SN 1993J we have $T_{\rm Ad,s} \sim 4.4~\rm{days} ~(t/t_0)$. The maximum energy fixed by balancing the acceleration and adiabatic loss timescales is
\beq
E_{\rm max,adi}\simeq [3.8\times 10^{-10} V_0~\rm{PeV}]~ \dot{M}_{-5}^{1/2} V_{\rm w,10}^{1/2} {\varpi \over \eta} \times \left({t \over t_0}\right)^{2m-1-m{s\over2}} \ .
\eeq

Again if the background magnetic field is $B_{\rm sat, NRS}$ then 
the maximum energies limited by losses have to be multiplied by ${\cal A}$, then $E_{\rm max,nuc} \propto (t/t_0)^{3(m-1)+ms/2}$, and $E_{\rm max, Adi} \propto (t/t_0)^{3m-2-ms/2}$.

\subsection{Time dependent maximum CR energy}\label{S:EMT}
\begin{figure}
\centering
\includegraphics[width=9cm]{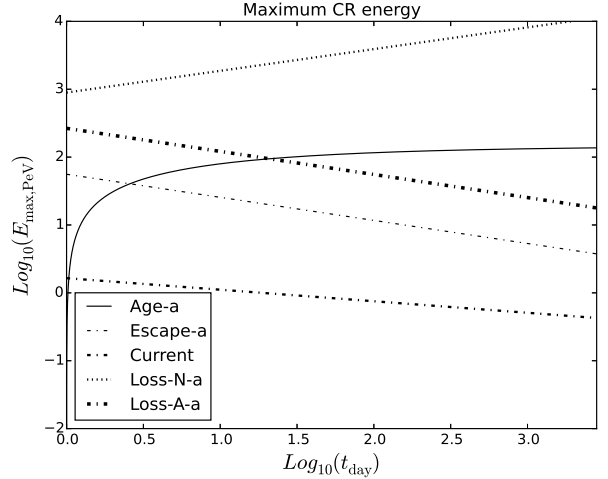}
\caption{ Maximum CR energy limits in PeV units for the model P as a function of time after shock breakout for the fiducial case of SN 1993J if the background field has been amplified up to $B_{\rm sat, NRS}$. The dotted line plots $E_{\rm max, nuc}(t)$, the large dot-dashed line plots $E_{\rm max, adi}(t)$, the intermediate dot-dashed line plots $E_{\rm max, cur}(t)$, the small dot-dashed line plot $E_{\rm max,esc}(t)$, the solid line plots $E_{\rm max,age}(t)$ The following parameters have been used: $\varpi=1$, $\eta=1$, ${\cal N}=5$, $\phi=14$, $\bar{\sigma}_{\rm pp}$=1.87.}
\label{F:EMA}
\end{figure}

\begin{figure}
\centering
\includegraphics[width=9cm]{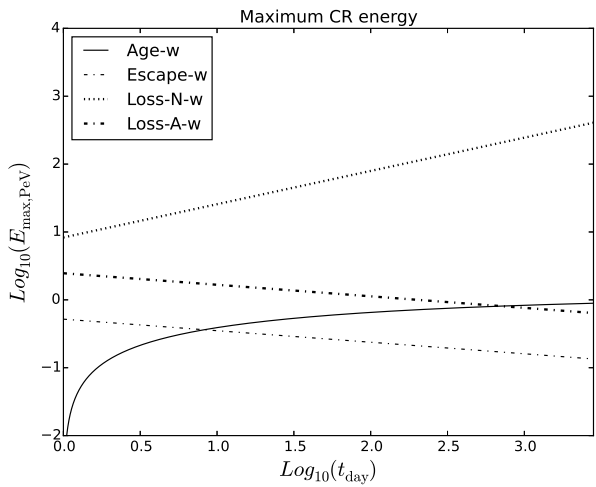}
\caption{ Maximum CR energy limits in PeV units for the model P as a function of time after shock breakout for the fiducial case of SN 1993J if the background field is $B_{\rm w}$. The dotted line plots $E_{\rm max, nuc}(t)$, the large dot-dashed line plots $E_{\rm max, adi}(t)$,  the small dot-dashed line plots $E_{\rm max,esc}(t)$, the solid line plots $E_{\rm max,age}(t)$. The following parameters have been used: $\varpi=1$, $\eta=1$, $\bar{\sigma}_{\rm pp}$=1.87.}
\label{F:EBW}
\end{figure}

In figure \ref{F:EMA} the maximum CR energy limits $E_{\rm max, age},E_{\rm max, esc},E_{\rm max, cur}, E_{\rm max, nuc}$ and $E_{\rm max, adi}$ are shown for model P as a function of time after the shock outburst, for the case when the NRS instability has time to amplify the background magnetic field to $B_{\rm NRS, sat}$. Note that T09 considered this case case only and omitted the effect of pp and adiabatic losses. Figure \ref{F:EBW} shows the maximum CR energy limits $E_{\rm max, age},E_{\rm max, esc}, E_{\rm max, nuc}$ and $E_{\rm max, adi}$ in the case when the background magnetic field is assumed to be the wind field.

At a given time $t$ the maximum CR energy is $E_{\rm max}(t) = \rm{Min}(E_{\rm max, age},E_{\rm max, esc},E_{\rm max, cur}, E_{\rm max, nuc}, E_{\rm max, adi})$. If the NRS instability is active in the CR precursor, as soon as $E_{\rm max,cur} < E_{\rm max, age-a}$ then $E_{\rm max} = E_{\rm max, cur}$. This occurs at $t > 1.1 t_0$ in figure \ref{F:EMA}, and the peak value for $E_{\rm max} \sim 2$ PeV which then drops as $t^{-0.17}$. At $t=10$ days ($t=100$ days), we have $E_{\rm max} \sim 1.4$ PeV ($E_{\rm max} \sim 0.9 $ PeV). We find $E_{\rm max} \sim 680$ TeV after one year, which is comparable, within a factor of 2, with values derived by \citet{Schure13} for SN shocks propagating in RSG winds. Note that adiabatic losses dominate over the losses due to pp collisions. If the CR current is not strong enough to generate a strong magnetic field amplification then $E_{\rm max}= E_{\rm max, esc-w}$. This occurs at $t > 12~t_0$ in figure \ref{F:EBW} where have $E_{\rm max} \sim 300$ TeV. Then $E_{\rm max}$ drops as $t^{-0.17}$ to $\sim 110$ TeV after one year. \\
For SN 1986J, at $t=5$ days we find $E_{\rm max} \sim 6$ PeV. Again for the reasons addressed in section \ref{S:CRG}, this number has to be considered as an upper limit. For SN 2009bb, at $t=20$ days we find $E_{\rm max} \sim 2.5$ PeV.

\section{Discussion} \label{S:DIS}
In this study we show that SNe can produce particles up to multi-PeV energies 
via the combination of fast shocks (velocity of order 0.1c), a high density CSM produced by stellar winds, and low wind magnetizations. A high degree of CSM
ionization can ease the particle acceleration process, but cannot be
assumed and in general is unlikely. Assuming that the background magnetic field
has a turbulent component, different instabilities driven by the
acceleration process can grow over intra-day timescales. This model is applied to
a set of powerful SNe (both type II and type Ib/Ic) detected at radio wavebands by the VLA and by VLBI.

This first study should be seen as a proof of concept. A full derivation of the time dependent CR distribution and gamma-ray emission is required to obtain a testable model. Parameters affecting the early gamma-ray emission from SNe include the
ratio of the mass-loss rate to the wind velocity ($\dot{M}/V_{w}$),
which determines the CSM medium density and thereby affects the CR
driven instability growth rate. The shock velocity controls the growth
rate of the instabilities and the acceleration timescale. The degree
of ionization is important for the particle acceleration efficiency,
and may also produce element dependent CR spectra in the case of
partial ionization. The background magnetic field is partly responsible for the
local magnetization and the shock obliquity. The SN luminosity affects
the gamma-gamma absorption process. In case the CR pressure becomes
larger than 10\% of the shock ram pressure non-linear calculations are mandatory
to find the final particle and photon distributions. Finally the gamma-ray detectability should be restricted to nearby events. The preliminary estimations made in \citet{Marcowith14} show an horizon of detectability at 1 TeV for the Cherenkov Telescope Array of $\sim 10$ Mpc.

Only about 5-6\% of the local core-collapse SNe have been classified
as Type IIb such as SN 1993J \citep{Smartt09}. There is a further
subdivision into two classes of IIb SNe, with compact and extended
progenitors \citep{Chevalier10}. The ones with extended radii, such as SN~1993J,
seem to have higher mass-loss rates and lower wind velocities, and are
the more promising candidates for high energy cosmic ray acceleration and the early detection of $\gamma$-ray emission due to their higher wind density.  The compact ones have wind velocities similar to those of Wolf-rayet stars, and thus
correspondingly lower wind densities. Therefore they may not be likely
candidates for detecting early $\gamma$-ray emission, thus further
reducing the observable sample. 

Type IIn SNe are probably the most promising targets for gamma-ray
telescopes in terms of high ambient density without significantly
reduced velocities. In the case of the Type IIn SN 1996cr, it has been
deduced, using numerical simulations that managed to reproduce the
X-ray spectra over more than a decade of evolution, that the shock was
interacting with a shell of density $\sim 10^5\, {\rm cm}^{-3}$
\citep{Dwarkadas10} a few years after explosion. This provides a high
density target for producing $\gamma$-rays via pion production, and is
the basis for taking IIns as promising candidates for early
$\gamma$-ray emission. These densities however are still lower than
those suggested by some authors \citep{Murase11, Murase14} in their
calculations. The latter calculations suggest very high $\gamma$-ray
fluxes. Type IIn SNe
would, for the same reasons, also be promising targets for detecting
neutrino emission from secondaries. Unfortunately the number of Type IIn  SNe is pretty small, comprising less than 4\% of the total core-collapse
population. It is
likely that numerical simulations would need to be done to effectively
deal with the SN hydrodynamics in the ambient medium. We will consider
Type IIn SNe in a later paper.

SNe IIP comprise the largest class of core-collapse SNe, making up
around half the total. Their progenitors are RSG stars, which have
wind mass-loss rates ranging from $10^{-7}$ to $10^{-4}$ M$_{\odot}$
yr$^{-1}$ \citep{Mauron11}. However, \citet{Smartt09} demonstrated that
observed progenitors of Type IIP SNe all appear to have masses below
about 16.5 $\msun$. Similarly \citet{Dwarkadas14} showed that IIPs have
the lowest X-ray luminosities amongst all core-collpase SNe, and thus
put an upper limit of 19 M$_{\odot}$ on the initial mass of their
progenitors, with correspondingly lower mass-loss rates
\citep{Mauron11}. The low mass-loss rates will result in lower maximum
energies than calculated for SN 1993J (see \citet{Cardillo15}).

The rare Ib/Ic SNe harbor the fastest shock waves. They are assumed to
arise from Wolf-Rayet progenitors, which have wind velocities two
orders of magnitude greater than RSGs, and therefore should have
corresponding wind densities two orders of magnitude lower. The fast
shock velocities are consistent with the lower densities. These shocks
tend to accelerate particles more efficiently to higher energies, and
their X-ray flux is presumed to be due to Inverse Compton or
synchrotron emission \citep{Chevalier06}, suggesting accelerated
electrons. It is possible therefore that the shocks are capable
of accelerating protons to high energies although our analysis suggests that due to lower wind densities CR instability growth rates can be reduced in such type of objects. Some W-R stars are surrounded by low density wind-blown bubbles bordered by a high density shell. If the shell is formed soon before the explosion, as is the case for the SN 2006jc \citep{Foley07}, then it provides a good target for accelerated protons to collide with. Such W-R stars may be good candidates for the early detection of gamma-ray emission. 

There may exist SNe similar to SN 1987A, whose progenitor, a
blue-supergiant, had a very low wind mass-loss rate wind on order
10$^{-8}$ $\msun$ yr$^{-1}$\citep{Chevalier95}, but which shows evidence for
a dense HII region with density of order 200 particles cm$^{-3}$
\citep{Dewey12}, surrounded by a dense circumstellar ring with density
$\sim$ 10$^4$ particles cm$^{-3}$ at a distance of $\sim 0.2$ pc from
the SN. Finally, other promising targets would include the class of
super-luminous SNe, especially those that are H-rich
\citep{Nicholl15}, as these may be interacting with extremely dense
environments. High densities close in to the star could favor fast instability 
growths and could again provide
target material for proton-proton collisions and detectable
$\gamma$-ray emission at an early age. 

Finally, we mention here the extremely rare Type Ia-CSM class, which appear to have the highest ambient densities as a class. It is not
clear what the progenitors of these SNe are, and in
fact whether they are bona-fide Ia's, but densities inferred
for the surrounding medium are as high as 10$^8$ cm$^{-3}$
\citep{Deng04, Aldering06, Bochenek18}, at least in the first couple of
years. If a close SN of this type were detected, they would be perhaps
the most likely to show detectable gamma-ray emission. Unfortunately,
these are the rarest class, and much like the IIns, the density
structure is not well known but expected to be quite complex, with
perhaps a two-component surrounding medium, not amenable
to analytic calculations, and thus require detailed
modeling.

\section{Conclusions} \label{S:CON}
The main conclusions of the study are as follows:
\begin{itemize}

\item Magnetic field strengths inferred from radio observations of a sample of powerful and nearby
type II and type Ib/Ic SNe can be, at least partly, explained by the process of magnetic field amplification
driven by CR driven instabilities.

\item We find that in fast shocks moving in dense CSM as it is the case of many type II SNe, the non-resonant streaming instability
can develop in the shock precursor in parallel shock configuration. Saturated magnetic field strengths can reach
up to $\sim$ 50\% of the magnetic field deduced from the modeling of radio lightcurves. This number accounts for 
both magnetic field amplification in the CR precursor and the transversal magnetic component compression at the shock front. Perpendicular shocks
may also trigger CR driven instabilities but only in the configuration of a sub-equipartition wind magnetic field energy density with respect to the wind kinetic 
energy density.

\item In our sample we find that SN 1986J and SN 1993J were the most efficient at generating turbulent magnetic fields and accelerating
cosmic rays. We find that for these cases maximum CR energies reach $\sim$ 1-10 PeV within a few days after the explosion and $\sim$ 0.1-1 PeV after one year. The upper limits can shift to higher energies if long wavelength magnetic perturbations can be generated. So Type IIn or compact Type IIb SNe may be good candidates for the Pevatron CR class. However, an accurate shock dynamics modelling of type IIn SN requires to go beyond the self-similar solution adopted in this study. The results obtained for SN 1986J have to be taken as upper limits. Finally we find that the trans-relativistic SN 2009bb can accelerate CR at energies in the range 2-3 PeV within a few days after the outburst.

\end{itemize}

In a subsequent paper, we will include a detailed calculation of
gamma-gamma opacity. This will allow us to derive a
time-dependent gamma-ray flux in the TeV domain, and to make accurate
predictions for gamma-ray detectability of SNe with HESS and the
future Cherenkov Telescope Array (CTA). This modeling will also include
the multi-wavelength emission produced by secondary particles as a
result of charged pion decay. We will calculate the expected high-energy neutrino flux from these SNe and compare it to the flux sensitivity of current and future neutrino facilities. A final study will treat SN IIn separately, as the wind density profile for these objects is more
complicated and require more refined treatment than the self-similar calculations used in this study. 
These objects require numerical modeling in order to derive shock dynamics
and evaluate particle acceleration and multi-messenger emission
efficiencies.\\

{\bf Acknowledgments:} This research collaboration is supported by a grant from the FACCTS program to the University of Chicago (PI: VVD; Co-I: MR, Univ of Montpellier). We are grateful to this program for funding travel between Chicago and Montpellier for VVD and MR. This work is supported by the ANR-14-CE33-0019 MACH project. AM thanks P.Blasi, A.Bykov, L.Dessart, D.Ellison for helpful discussions.

\bibliographystyle{mnras}
\bibliography{Marco} 



\end{document}